\documentclass[aps,cyrwin,prb,twocolumn,floats,eqsecnum,graphicx,epsfig,euscript,amsmath,amssymb]{revtex4}
\usepackage{mathrsfs}
\usepackage[dvips]{graphicx}
\usepackage{epsfig}
\DeclareMathAlphabet\EuScript{U}{eus}{m}{n} \SetMathAlphabet\EuScript{bold}{U}{eus}{b}{n}

\def\lapprox{\,\raise0.4ex\hbox{$<$}\kern-0.8em\lower0.7ex\hbox{$\sim$}\,}
\def\gapprox{\,\raise0.4ex\hbox{$>$}\kern-0.8em\lower0.7ex\hbox{$\sim$}\,}
\def\be{\begin{equation}}
\def\ee{\end{equation}}
\def\ba{\begin{eqnarray}}
\def\ea{\end{eqnarray}}

\begin{document}
\title
{Spin-rotation mode in a quantum Hall ferromagnet }

\author{$\qquad$ S. Dickmann}

\affiliation{$$Institute of Solid State Physics, Russian
Academy of Sciences, Chernogolovka, 142432 Russia}

\date{\today}

\begin{abstract}
A spin-rotation mode emerging in a quantum Hall ferromagnet due to laser pulse excitation is studied. This
state, macroscopically representing a rotation of the entire electron spin-system to a certain angle,
is not microscopically equivalent to a coherent turn of all spins as a single-whole and is presented in the form of a combination of eigen quantum states corresponding to all possible $S_z$ spin numbers. The motion of the macroscopic quantum state is
studied microscopically by solving a non-stationary Schr\"odinger equation and by means of a kinetic
approach where damping of the spin-rotation mode is related to an elementary process, namely, transformation of a `Goldstone spin exciton' to a `spin-wave
exciton'. The system exhibits a spin stochastization mechanism (determined by spatial fluctuations of the Land\'e
factor) ensuring damping, transverse spin relaxation, but irrelevant to decay of spin-wave
excitons and thus not involving longitudinal relaxation, i.e.,recovery of the $S_z$ number to its equilibrium
value.\vskip 1mm

\end{abstract}
\maketitle

\section{Introduction (macroscopic approach)}

\vspace{-2mm}

Two-dimensional  electron gas (2DEG) composed only of conduction-band electrons embedded in quantized perpendicular or tilted
magnetic field represents a unique quantum object for direct study of magnetic phenomena and collective spin
excitations using both macroscopic and microscopic approaches. In particular, in the so-called quantum Hall
ferromagnet (QHF), i.e. in the case of a large nonzero total spin momentum (i.e. at fillings $\nu\!=\!1,3,...$ or even at
$\nu\!=\!1/3,1/5...$), it is possible only by means of free conduction-band electrons to experimentally model and study
properties inherent to common exchange magnets.\cite{pi92,ba95,ku05,va06,ga08,dr10,wurst,Fukuoka,la15,zh14} Many QHF
properties (for example, spectra of magnetoplasma and spin excitations as well as spectra of spin-magnetoplasma
excitations$\,$\cite{pi92,va06,ga08,dr10,theory}) are determined directly by the `ab initio' interaction, Coulomb coupling
of 2D electrons. Besides, external fields such as spatial electrostatic fluctuations within the 2D structure and
spin-affecting microscopic couplings, actually spin-orbit and hyper-fine ones, both responsible for the dephasing
and relaxation processes, are also considered straightforwardly in the context of a perturbative approach. The QHF
features are substantially different from description of ordinary magnets, e.g., with spatially fixed spin positions, which usually represents a phenomenological approach or a microscopic study based on a model Hamiltonian.

Description of dynamics of the ferromagnet by means of the Landau-Lifshitz (L-L) equation\cite{ll} is just a typical
phenomenological approach. In fact, this well-known equation { consistent with general principles} {is not even
derived but just proposed}. In the case relevant to the QHF it would be: $\hbar\partial{\bold S}/\partial
t\!=\!-g\mu_{\rm B}{\bold S}\!\times\!{\bold B}\!-\!\lambda{\bold S}\!\times\!({\bold S}\!\times\!{\bold B})$, where
${\bold S}$ is a macroscopically large electron spin. (It is taken into account that the effective magnetic field in
rarefied electron gas is equal to external magnetic field ${\bold B}$.) The first term in the RHS of the
L-L equation is proportional to the magnetic moment of the spin and determines the fast precession process around
${\bold B}$ with frequency $|g|\mu_{\rm B}B/\hbar$. This term is definitely valid also in the QHF case. The second term,
according to the authors\,\cite{ll} should be a relativistic correction responsible for precession damping,
hence, describing a slow approach from ${\bold S}$ to ${\bold B}$. This term is chosen in the form corresponding to
the spin motion conserving length of ${\bold S}$ ($\partial{\bold S}^2\!/\partial t\!\equiv\!0$), i.e. a variation of
the absolute value of ${\bold S}$ is disregarded. Such a conservation condition is natural for the strong exchange
ferromagnet where damping is accompanied by weak dissipative processes (in particular, by dissipation of Zeeman
energy $|g|\mu_{\rm B}B|S_z(\infty)\!-\!S_z(0)|$ due to restoration of the $S_z(t)$ component; $\!{\hat
z}\!\parallel\!{\bold B}$), yet not violating conservation of the total exchange energy considered to be strictly
determined by $S$.

It is worth noting that the characteristic exchange energy in the 2DEG is at least by two orders smaller than in ordinary `insulating'magnets, so the term {\em ferromagnet} as applied to the magnetized 2DEG is fairly conventional. In the QHF the
Coulomb/exchange interaction energy ($\sim\!E_{\rm C}=\alpha e^2/\kappa l_B$, where $\alpha\!<\!1$ is a form-factor
arising  owing to finiteness of the 2D layer thickness; $\kappa$ and $l_B$ are the dielectric constant and magnetic
length) undoubtedly represents the main force holding the electron spins aligned along the magnetic field. This fact is
manifested, for instance, in a gigantic increase in the effective $g$-factor obtained in measurements of activated
conductivity;\cite{usher} however, the absence of spontaneous magnetization in the 2DEG when the external magnetic field is
switched off, certainly indicates that the QHF is not an ordinary ferromagnet. Experimental
research $\,$\cite{Fukuoka,la15} and the microscopic study presented in the following sections show that under quantum
Hall conditions 2DEG spin-precession damping occurs via dephasing/stochastization processes not affecting the
exchange energy, while $S$ is still diminishing in accordance with the condition of constancy of the $S_z$ component
that corresponds to the Zeeman energy conservation. The subsequent process of Zeeman energy dissipation is related to the spin-wave relaxation/annihilation and  proceeds much slower. {It is indeed determined not only by thermal and spatial fluctuations responsible for energy dissipation but also by weak couplings, for instance by spin-orbit and hyper-fine couplings, responsible for the change of the $S_z$ component.  [See the theoretical estimates given in Ref. \onlinecite{di12} and references therein, and Ref. \onlinecite{zh14} presenting experimental measurement of  $S_z$ recovery (within time $\sim100-150\,$ns).]} Therefore, the total magnetic  relaxation in the QHF case is characterized by two stages: the first one, being comparatively fast, is actually damping of the spin precession where
the direction of ${\bold S}$ approaches the ${\bold B}$ direction at the $S_z$ held constant; the second stage related to the
Zeeman energy dissipation represents slow recovery of the spin angular momentum ${\bold S}$ (directed parallel to
${\hat z}$; $S\!=\!S_z$) to its equilibrium value.\cite{zh14}  In terms of nuclear magnetic resonance,\cite{abr} the characteristic times of these two stages could be called  transverse time $T_2$ for the
fast stage and longitudinal time $T_1$ for the slow stage.

Similar to the Landau-Lifshitz equation and on the basis of similar phenomenological ideas we can write out an
equation describing QHF spin motion in the framework of the macroscopic approach. Again the term responsible for
the precession damping is assumed to be a small correction proportional to a vector directed from ${\bold S}$ to
${\bold B}$. However, now, in accordance with the above requiring constancy of the $S_z$ component, the motion equation
should be written in the following simplest form
\begin{equation}\label{1.1}
\partial{\bold S}/\partial t\!=\!-(g\mu_{\rm B}\!/\hbar){\bold S}\!\times\!{\bold B}\!-\!\lambda_{\rm QH}{\bold B}\!\times\!({\bold S}\!\times\!{ \bold B}),
\end{equation}
where, contrary to the L-L equation, any variation of the $S_z$ component is disregarded. Constant $\lambda_{\rm QH}$ in Eq. \eqref{1.1} can only be found within a specific microscopic model studied in the following sections. For the $S_z$ and ${\bold
S}\!{}_{\perp}\!\!=\!(S_x,S_y)$ components we obtain: $\partial S_z/\partial t\!=\!0$
(instead of $\partial{\bold S}^2\!/\partial t\!\equiv\!0$ in the L-L equation) and
\begin{equation}\label{1.2}
\partial{\bold S}\!{}_{\perp}/\partial t\!=\!-(g\mu_{\rm B}\!/\hbar){\bold S}\!{}_{\perp}\!\!\!\times\!{\bold B}\!-\!\lambda_{\rm QH}B^2{\bold S}\!{}_{\perp}.
\end{equation}
The transverse relaxation time $T_2\!=\!1/\lambda_{\rm QH}B^2$ must be much larger than
the precession period $\hbar/g\mu_{\rm B}B$, i.e. we have necessary condition $\lambda_{\rm QH}\hbar B/g\mu_{\rm B}\ll 1$.
In Fig. 1b the trajectories of the ${\bold S}$ vector approaching the ${\hat z}$ direction are drawn in both situations:
the motion is ruled by the Landau-Lifshitz equation and by Eq. \eqref{1.1}. \vspace{-2mm} \vspace{3.8mm}
\begin{figure}[h]
\begin{center}
\vspace{-26.mm}
\hspace{-17.mm}
\includegraphics*[angle=0,width=.78\textwidth]{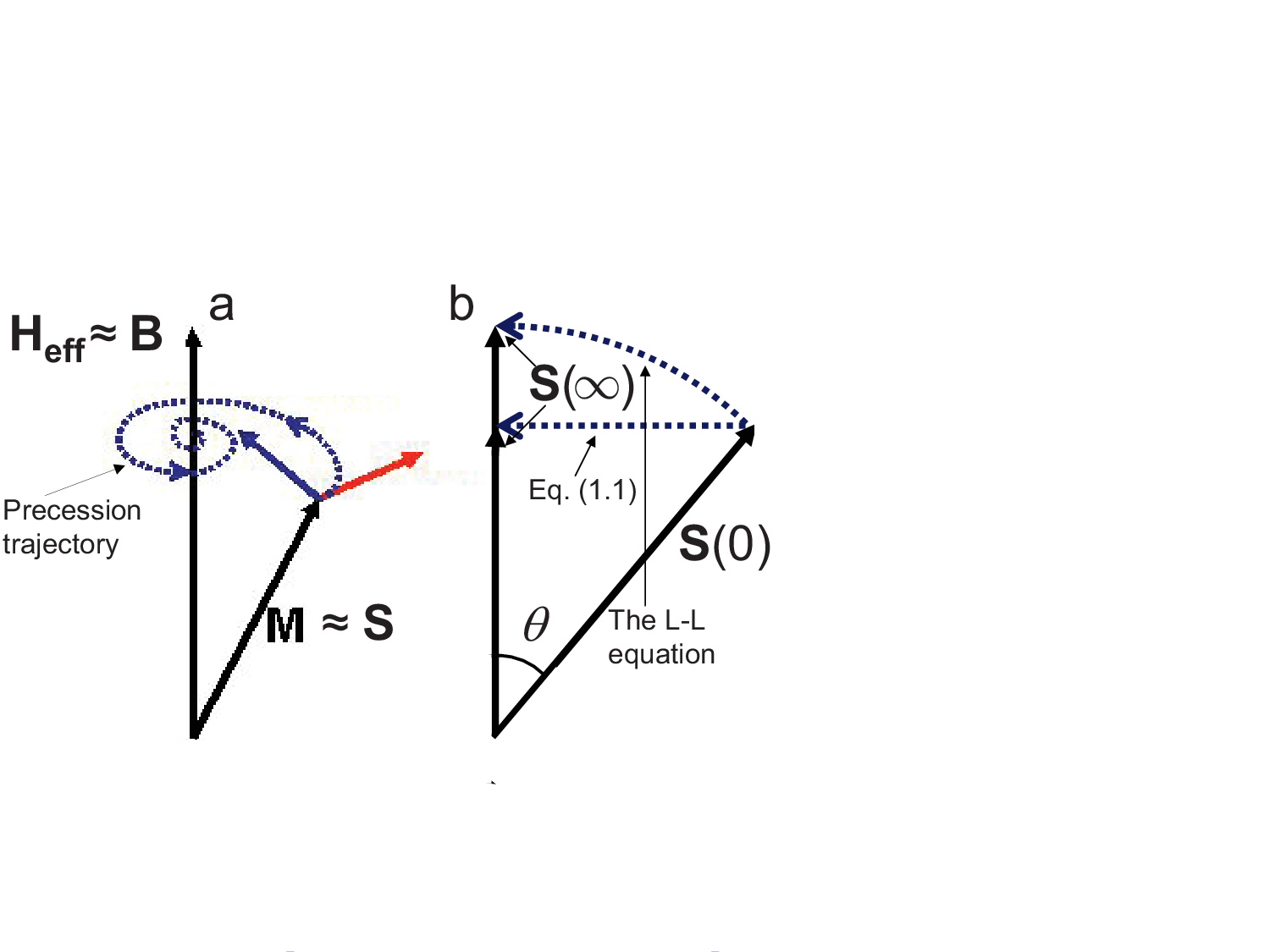}
\end{center}
\vspace{-17.mm}
 \caption{ a --- illustration of magnetic moment motion; the red arrow indicates the first (precession) term in
 the macroscopic equation, the blue arrow to the second (damping) term. b --- illustration of damping considered
 in the coordinate system precessing with spin momentum;  the blue dashed lines are the ${\bf S}(t)$ damping trajectories
 from ${\bf S}(0)$ to ${\bf S}(\infty)$ drawn for the Landau-Lifshitz equation (arc) and for Eq. (1.1) (horizontal line).}
\end{figure}

So, at the initial moment the spin-rotation mode is a macroscopic vector ${\bf S}(0)$ rotating by angle $\theta$
about an axis lying in plane $(\hat{x},\hat{y})$. Here $\theta$ measures deviation from the ground-state
magnetization direction $\hat{z}$ (see Fig. 1b).  If $0\!<\!\theta\!<\!\pi$, then rotation of ${\bf S}$ by any angle
$0\!<\!\varphi\!<\!2\pi$ about the $\hat{z}$ axis leads certainly to a different state but with the same
energy. Rotational symmetry, about ${\hat z}$, of the QHF state at any $0\!<\!\theta\!<\!\pi$ corresponds to group
{\bf C}${}_{1v}$ in the case and thus represents spontaneous breaking of the ground-state continuous symmetry {\bf C}${}_{\infty v}$.
This `$\theta$-inclined state' possesses energy $\epsilon_{\rm Z}(1\!-\!\cos{\theta})S(0)$ macroscopically corresponding
to a gapless Goldstone mode in terms of parameter $\theta$ ($\epsilon_{\rm Z}\!=|g|\mu_B B$). We will use the term `Goldstone mode' for $\theta$--spin-rotational
deviation in order to distinguish it from another one corresponding to a `longitudinal' deviation where both
spin numbers $S$ and $S_z$ change equally: $\delta S_z\!=\!\delta S$. It is obvious that in the latter case the symmetry of the
system remains {\bf C}${}_{\infty v}$ as in the ground state.

The main purpose of the present work is to study transverse relaxation, i.e. stochastization of the Goldstone mode,
and thus calculate inverse time $1/T_2\!=\!\lambda_{\rm QH}B^2$. In comparison with the estimate made in Ref.
\onlinecite{la15} we consider not only small but arbitrary deviation angles of spin ${\bold S}$ from its equilibrium
direction.

The calculation is performed within a microscopic approach.  As the initial non-equilibrium quantum state presenting spin rotation we study the combination\vspace{-1mm}
\be
\label{spin-rot}
|C\rangle = \sum_n C_n(S_-)^n|0\rangle\,, \vspace{-4mm}
\ee
where $|0\rangle$ is the QHF ground state and $S_-\!=\!{\hat S}_x\!-\!i{\hat S}_y$ the `spin lowering' operator of the whole system. The specific set of coefficients $\{C_n\}$ is determined by the prehistory of appearance of state \eqref{spin-rot} in the system. However, at any set $\{C_n\}$ this state has obviously the following properties: (i) the $|C\rangle$
vector is diagonal for the ${\bf S}^2$ operator corresponding to its maximum value in the ground state $|0\rangle$; (ii) any $|C\rangle$ are orbitally equivalent to the ground state: the matrix elements of any spin-independent fields, including Coulomb coupling, calculated within the brackets $\langle C|...|C\rangle$ are equal to those
calculated within $\langle 0|...|0\rangle$. In other words, the $|C\rangle$
state represents a kind of `${\vec k}=0$' excitation not disturbing the orbital state of the electron system. In Secs. II and III we explain in more detail our choice of the
initial quantum state and present coefficients $\{C_n\}$ correponding to the laser-induced spin-rotation mode \eqref{spin-rot}.

The stochastization mechanism considered here
is determined by smooth spatial fluctuations of the $g$-factor in the 2DEG, and has a simple physical meaning: within a
single-electron approach, the electrons do not precess coherently but with slightly different Larmor frequencies in
different places of the 2D space.\cite{ku17} It is interesting that, as in the case of the L-L equation, the damping
term in Eqs. \eqref{1.1} and \eqref{1.2} formally represents a relativistic correction, since the small ratio
$\hbar/g\mu_{\rm B}T_2$ turns out to be proportional to $c^{-1}$ [see Eqs. \eqref{T_2}, \eqref{T_2gauss} and
\eqref{T_2lorentz} below]. Besides, the studied below microscopic process destroying the spin-rotation mode enables to
find the damping rate proportional to the density of spin-wave states (purely electronic ones!). The density of states for
its part is inversely proportional to Coulomb coupling strength $E_{\rm C}$, i.e. {\em stronger coupling means
longer damping time}. In our case the coefficient $\lambda_{\rm QH}$ may be written as $\lambda_{\rm
QH}\!=\!\lambda'/r_s$ where the parameter $r_s$ represents the ratio of the Coulomb/exchange interaction energy to the
characteristic single-electron energy. For the QHF the former is $E_{\rm C}$ and the latter is cyclotron energy
$\hbar\omega_c$, so that $\lambda'$ proves to be independent of magnetic field for the stochastization
mechanism in question. Under typical quantum Hall conditions we have $r_s\!\sim\!0.2$. [For comparison: in the ordinary exchange
ferromagnet the parameter $r_s\!$ is huge, $\sim\!100\!-\!1000$, and the second term in Eqs. \eqref{1.1} and
\eqref{1.2} becomes negligible, so that the L-L equation is used in the case.\cite{foot}]

{It should be mentioned that other mechanisms of relaxation of the spin-rotation mode \eqref{spin-rot} were theoretically considered long before direct measurements of the relaxation rate in Refs. \onlinecite{Fukuoka} and \onlinecite{la15}. Works \onlinecite{di96} and \onlinecite{di04} were devoted to study of relaxation of the $(S_-)^n|0\rangle$ component, i.e., the case where
set $\{C_n\}$ consists of a single number $C_n$ was considered. The relaxation, -- stochastization of the Goldstone mode, -- was assumed to be related to spin-orbit Dresselhaus and Rashba couplings responsible for the change of the spin state in the presence of energy dissipation due to electron-phonon coupling$\,$\cite{di96} or the electrostatic interaction of an electron with an external random potential$\,$\cite{di04}. The calculated relaxation times were found to be much longer (in fact $>100\,$ns) than those measured later. In Ref. \onlinecite{by99}
the authors considered another type of state \eqref{spin-rot} (see below a `conventional' spin rotation mode) and an electron-spin---phonon relaxation mechanism which is even weaker than that studied in Refs. \onlinecite{di96} and \onlinecite{di04} and thus resulting in slower relaxation (see comment$\,$\cite{kh01}). So, all the three relaxation mechanisms$\,$\cite{di96,di04,by99,kh01} are irrelevant to the actual experimental results.\cite{Fukuoka,la15}}

It is significant that in the case of `classical' QHFs when fillings are odd-integer ($\nu\!=\!1,\,3...$), the
microscopic approach presented in the following sections enables us to solve the problem in an asymptotically exact
way in the case the parameter $r_s$ is considered to be small. The experimental data and theoretical discussion show
that exactly such `odd-integer' QHFs are the strongest, i.e. the precession damping is much longer compared to
nearby states with fractional fillings.\cite{la15}  Besides, the microscopic research allows
finding not only coefficient $\lambda_{\rm QH}$ but reveals the ${\bold S}\!{}_\perp\!(t)$ behavior, which is
absolutely beyond the macroscopic approach: in addition to the exponential damping governed by Eq. \eqref{1.2}, the
microscopic study shows that at short times $t\!\lesssim\tau_0\!\ll\!T_2$ there occurs an initial transient stage which is not
described by Eq. \eqref{1.2} . The $\tau_0$ value will be calculated in Sec. V.
\vspace{-3mm}

\section{Microscopic description of the system: the Hamiltonian and relevant eigenstates}

In the absence of any {interaction mixing spatial and spin variables}, the Hamiltonian of a translationally
invariant quantum-Hall system has the following form:
\vspace{-2mm}
\begin{equation}\label{Hamiltonian}
{\hat H}_0=-\epsilon_{\rm Z}{\hat S}_z+{\hat H}^{(1)}+{\hat H}_{\rm Coul}.\vspace{-1mm}
\end{equation}
Here ${\hat S}_z\!=\!\frac{1}{2}\sum_i\sigma_{iz}$ ($\sigma_{iz}$ is the Pauli matrix). The `kinetic energy' electron operator and the Coulomb-interaction operator,
\vspace{-2mm}
\begin{equation}\label{operators}
\begin{array}{l}
\displaystyle{{\hat H}\!{}^{(1)}\!\!=\!\sum_i\!\frac{({\hat {\bf p}}_i+e{\bf A}_i)^2}{2m^*}}\qquad\qquad\vspace{-1mm}\\\displaystyle{\qquad\qquad \mbox{and}\quad{\hat H}_{\rm Coul}\!=\frac{1}{2}\!\sum_{i\neq\!j}\!U({\bf R}_i\!-\!{\bf R}_j)},\vspace{-2mm}
\end{array}
\end{equation}
are those acting only on electron spatial variables. [{Here and everywhere below we set $\hbar\!=\!1$}; $i$
and $j$ are subscripts numbering electrons; ${\hat {\bf p}}_i$ is the 2D electron momentum operator ($m^*$ stands for
the electron effective mass); ${\bf R}_i$ is the 2D radius-vector in the quantum-well plane given by the $\{{\hat X},
{\hat Y}\}$ coordinate system not related to the 3D system $\{{\hat x},{\hat y},{\hat z}\}$ for the spin space; ${\bf
A}_i\equiv{\bf A}({\bf R}_i)=(0;B\!{}_{\perp}\!X_i)$ is the 2D vector-potential operator, where $B\!{}_\perp$ is the
component perpendicular to the 2DEG plane -- the latter may be tilted with respect to the ${\bold B}$ direction (${\bf
B}\!\parallel\!{\hat z}$).]\vspace{-4mm}

\subsection{`Spin-deviation' eigenstates with $S_z<S$}
\vspace{-2mm}

Any purely spin operator commutes with $H^{(1)}$ and $H_{\rm Coul}$. Thus, the spin lowering operator $S_-$ can
play the role of a generator of `spin-deviation' eigenstates. Indeed, let $|0\rangle$ be an eigenstate of the quantum
Hall system corresponding to exact spin quantum numbers equal to $S_z\!=\!S_{0z}$ and $S\!=\!S_0$, and to energy $E_0$.
Then the $S_-|0\rangle$ state is also an eigenstate. By acting with ${\hat S}_z$ and ${\bf S}^2\!\equiv\!{\hat
S}_x^2\!+\!{\hat S}_y^2\!+\!{\hat S}_z^2$ on this state one gets $(S_{0z}\!-\!1)S_-|0\rangle$ and
$S_0(S_0\!+1\!)S_-|0\rangle$, respectively. Operating with ${\hat H}_0$ on the $S_-|0\rangle$ state we obtain
$(\epsilon_{\rm Z}\!+\!E_0)S_-|0\rangle$. So, the action of the $S_-$ operator {\em does not change the total spin number
but only results in the $S_z\!\to\!S_z\!-\!1$ change}. Besides, this action does not affect the orbital state of the electron
system.

A set of states defined as\vspace{-2mm}
\begin{equation}\label{n-states}
|n\rangle=\left(S_-\right)^n|0\rangle\vspace{-2mm}
\end{equation}
represents exact eigenstates orbitally equivalent to state $|0\rangle$ but with spin numbers $S_z\!=\!S_{z0}\!-\!n$ and $S\!=\!S_{0}$, and energies\vspace{-1mm}
\begin{equation}\label{energy_n}
E_n=E_0\!+\!\epsilon_{\rm Z}n.\vspace{-1mm}
\end{equation}

It is worth to note that even if $n$ is macroscopically large {\em a single $|n\rangle$ state does not
describe any dynamics of the Goldstone mode because this stationary state has no definite azimuthal orientation.}
Indeed, the quantum-mechanical average of the ${\bf S}_\perp$ vector is vanishing: $\langle n|{\hat
S}_x|n\rangle\!=\!\langle n|{\hat S}_y|n\rangle\!\equiv\!0$ due to the obvious equality $\langle
n|S_-|n\rangle\!\equiv\!\langle n|n\!+\!1\rangle\!\equiv\!0$. Meanwhile at large $n$ if the spin
component $S_z$ in the $|0\rangle$ state takes the largest possible value, i.e., if $S_{z0}\!=S_0\equiv S$) the squared transverse
component in the $|n\rangle$ state may still be macroscopically significant: $S_\perp^2\!\approx\!S^2\!-(S\!-\!n)^2$;
hence, a macroscopic deviation angle appears: $\theta\!\approx\!\arcsin{(S_\perp\!/S)}\!=\!\sqrt{{n}(2\!-{n}/{S})\!/S}$.
(Here $n$ is considered to be $\gg\!1$ and, besides, $2S\!-\!n\!\gg\!1$.) Relaxation of a single state $|n\rangle$ may be studied, actually representing a key problem for study of relaxation of any state given by Eq. \eqref{spin-rot}. In the case of maximum $S_{z0}\,(=\!S_0)$, we have equality
$S_+|0\rangle\!\equiv\!0$ (here $S_+\!=\!S_-^\dag$)  which allows calculating squared norm $\langle
n|n\rangle$:\vspace{-1mm}
\begin{equation}\label{norm_n}
R_{2S,n}=\langle n|n\rangle=\frac{(2S)!n!}{(2S\!-n)!}R_0,\vspace{-1mm}
\end{equation}
where $R_0\!\equiv\!R_{2S,0}\!=\!\langle 0|0\rangle$. If $|0\rangle$ is the ground state, then in the specific case of the odd-integer quantum Hall ferromagnet the spin number $S_0$ is the maximum total spin of electrons completely occupying the spin-up sublevel of the
Landau level, therefore $2S_0$ is equal to Landau-level degeneracy ${\cal N}_\phi$.
If the $R_0\!\equiv\!1$ condition is chosen
[see Eq. \eqref{2.1} below], then
we find $\langle n|n\rangle=R_{{\cal N}_\phi,n}$, where\vspace{-1mm}
\begin{equation}\label{norm_Nphi_n}
R_{{\cal N}_\phi,n}=\frac{{\cal N}_\phi!n!}{({\cal N}_\phi\!-n)!}.\vspace{-2mm}
\end{equation}

Concluding this subsection, it should be noted that, if the number of terms in the combination $|C\rangle$ [see Eq. \eqref{spin-rot}] is larger than one, then $|C\rangle$
is not an eigenstate of the Hamiltonian \eqref{Hamiltonian}. Generally, for arbitrary combination $|C\rangle$ of
eigenstates there is no direction ${\hat z}'$ in spin space where the spin projection $S_{z'}$ would be an eigen
quantum number. (Such states are called states with partial spin polarization of particles;\cite{LL3} the only exception to this general situation is a
special case when all electron spins are equally aligned along axis ${\hat z}'$ inclined by a definite angle $\beta$ to
the ${\bf B}$ direction.\cite{con-spin-rot}) Now, however, the quantum average of transverse spin ${\bf S}_\perp$ is not
equal to zero and is completely determined by the $\{C_n\}$ set. Taking into account that\vspace{-1mm}
 \begin{equation}\label{orthogonality}
\langle n|S_+|m\rangle=\delta_{m,n\!+\!1}R_{{\cal N}_\phi,m}\vspace{-1mm}
 \end{equation}
($\delta_{...}$ is the Kronecker delta) and calculating $\langle C|S_+|C\rangle=\sum_n\!C_n^*C_{n\!+\!1}\langle
n\!+\!1|n\!+\!1\rangle$, we find the values of components $\langle S_x\rangle=\mbox{Re}\langle C|S_+|C\rangle$ and
$\langle S_y\rangle=\mbox{Im}\langle C|S_+|C\rangle$.

So, the $|C\rangle$ state may be considered as a microscopic representation of the Goldstone mode whose subsequent
evolution is governed by the non-stationary Schr\"odinger equation. In the following, in order to emphasize the
role of elementary $S_-|0\rangle$ spin excitation in formation of the Goldstone mode, we call it {\em `Goldstone
spin exciton'} or simply {\em Goldstone exciton}. Spin-deviation state \eqref{n-states} formally represents
Goldstone exciton condensate provided $n$ is macroscopically large. \vspace{-3mm}

\subsection{`Spin-wave' eigenstates -- excitations corresponding to change of spin numbers:
$\delta S=\delta S_z=-1$}

Since the eighties it is known that in a translationally invariant QHF there are low-energy excitations -- {\em spin-wave
excitons} characterized by 2D momentum ${\bf q}\!\neq\!0$ (just like in an ordinary ferromagnet whose dynamics
is governed, e.g., by the Heisenberg Hamiltonian). At odd-integer filling such states and their energies may be
calculated within the leading approximation in $r_s$, which actually permits to use the single-Landau-level
approach.\cite{theory} So, considering, that the number of electrons in the $l$-th highest (nonempty) Landau level is
equal to Landau level degeneracy $N_e\!=\!{\cal N}_\phi$ and assuming that all lower levels are completely
occupied, as a ground state we have \vspace{-3mm}
\begin{equation}\label{2.1}
|{0}\rangle\!=\!|\overbrace{\uparrow\!\uparrow\!...\!\uparrow}^{\hphantom{N}
{}_{{}_{{}_{{\cal N}_\phi}}}}\,\rangle\equiv a^\dag_{p_1}a^\dag_{p_2}...a^\dag_{p_{{\cal N}\!\!{}_\phi}}|{\rm vac}\rangle,\vspace{-2mm}
\end{equation}
where $a_p^\dag$ is the operator creating a {\em spin-up} (along the ${\bf B}$) electron in the $p$-th state of the degenerate Landau level.
To define $|0\rangle$ uniquely, we consider the $p_\kappa\!\!=2\pi\kappa/L$ numbers in
Eq.$\;$\eqref{2.1} to be ordered by taking consecutive values $2\pi/L,\,4\pi/L,...\,2\pi\!{\cal N}_\phi/L\!\equiv\!L/l_B^2$ , where $L^2$ is the area of the 2D system. The Landau-level eigenfunctions are $\psi_p({\bf
R})\!\propto { H}_l(\frac{X\!-\!x_p}{l_B})\exp{[-\frac{(X\!-\!x_p)^2\!+2{i}x_pY}{2l_B^2}]}$, where
$\!x_p\!=\!p\,l_B^2\!$,
and $\!{ H}_l$ is the Hermite polynomial. The terms of the Coulomb interaction Hamiltonian
\eqref{operators} are presented in the secondary quantization form $\sum_{i,j}\!U({\bf R}_i\!-\!{\bf R}_j)\to
\int\!\!\int\! d^2R_id^2R_j{\hat \psi}^\dag({\bf R}_j){\hat \psi}^\dag({\bf R}_i)U({\bf R}_i\!\!-\!{\bf R}_j){\hat
\psi}({\bf R}_i){\hat \psi}({\bf R}_j)$. Here within the single Landau-level approximation we have ${\hat \psi}({\bf
R})=\sum_p\!\left({a_p\atop b_p}\right)\!\psi_p({\bf R})$, where $b_p$ is the {\em spin-down} electron annihilation
operator in the $p$-th state of the same $l$-th level. [Averaging over the quantum-well width for the $U({\bf R})$
Coulomb vertex {is assumed to be performed}.] This two-sublevel approach has been repeatedly
used$\,$\cite{dz83,ra86,dz91,by96,di-iord,di04,di12} (see also the relevant expressions for ${\hat H}^{(1)}$ and ${\hat
H}_{\rm Coul}'$ in Appendix A). It allows to describe the spin-wave excitation $|{\bf q};\!1\rangle\!\equiv\!{\cal
Q}_{\bf q}^\dag|0\rangle$ by means of a spin-exciton creation operator,\vspace{-1mm}
\begin{equation}\label{QQ}
{\cal Q}_{\bf q}^{\dag}=\sum_{p}
  e^{-iq_x\!p}\;
  b_{p+\frac{q_y}{2}}^{\dag}a_{p-\frac{q_y}{2}}.\vspace{-2mm}
\end{equation}
[Cf.$\;$also Refs.$\;$\onlinecite{la15} and \onlinecite{zh14} -- the previous definitions of the ${\cal Q}^\dag$-operators
differ from Eq. \eqref{QQ} by factor ${\cal N}_\phi^{-1/2}$; ${\bf q}$ and $p$ in Eq. \eqref{QQ} are measured in
$1/l_B$ units]. Energy of the $|{\bf q};\!1\rangle$ spin-wave exciton to the first order in the Coulomb interaction is found by the action of the reduced Coulomb-coupling operator: ${\hat H}_{\rm Coul}'|{\bf q};\!1\rangle=[{\hat H}_{\rm
Coul}',{\cal Q}_{\bf q}^{\dag}]|0\rangle\!+E_0'|{\bf q};\!1\rangle$, where $E_0'$ is the Coulomb part of ground state energy $E_0\!=\!E_0'\!+\!{\cal N}_\phi(\omega_c\!-\!\epsilon_{\rm Z})\!/2\,$ ($\omega_c\!=\!eB_\perp/m^*c$). Then (see Appendix A) we get the Coulomb part ${\cal E}_{q}$ of the spin-wave energy obtained to the first order in parameter $r_s$.\cite{theory} At small
momenta ${\bf q}$ (when $q\!\ll\!1$, meaning $ql_B\!\ll\!1$ in common units) the spectrum is quadratic: ${\cal E}_{
q}\!=\!q^2[1/2M_{\rm x}\!+\!O(r_s^2\omega_c)]$. The spin-wave exciton mass $M_{\rm x}$ was not only calculated but
experimentally measured;\cite{ga08} actually $M_{\rm x}^{-1}\!\sim\!2\,$meV in the typical wide-thickness GaAs/AlGaAs
quantum Hall systems.

The ${\hat S}_z$ spin operator in this representation takes the form ${\hat S}_z\!=\!\frac{1}{2}\sum_p(a_p^\dag
a_p\!-b_p^\dag b_p)$. As $[{\hat S}_z,{\cal Q}_{\bf q}^\dag]=-{\cal Q}_{\bf q}^\dag$, we get\vspace{-3mm}
\begin{equation}\label{Sz_action}
 {\hat S}_z|{\bf q};\!1\rangle=\left(\frac{{\cal N}_\phi}{2}-1\right)|{\bf q};\!1\rangle,\vspace{-2mm}
\end{equation}
that is the spin wave reduces the $S_z$ number by $\delta S_z\!=\!-1$. The operator ${\hat H}^{(1)}$ commutes with
${\cal Q}_{\bf q}^\dag$ (see Appendix A), hence, the energy of the spin-wave state found from the Schr\"odinger
equation ${\hat H}_0|{\bf q};1\rangle=E_{1,{\bf q}}|{\bf q};1\rangle$ is
\vspace{-1mm}
\begin{equation}\label{s-w_energy}
E_{1,{\bf q}}=\epsilon_{\rm Z}+{\cal E}_{\bf q}+E_0.\vspace{-1mm}
\end{equation}
The quantum average of the spin transverse component $S_\perp$ in the spin-wave state vanishes because $\langle 1,{\bf
q}|S_+|{\bf q};1\rangle\!\equiv\!0$.

Now we pay attention to operator equivalence \vspace{-2mm}
\begin{equation}\label{Q_equiv}
{\cal Q}_{\{{\bf q}\equiv{\bf 0}\!\}}^\dag\equiv S_-.\vspace{-1mm}
\end{equation}
In spite of this, the spin-wave exciton $|{\bf q};1\rangle\!=\!{\cal Q}_{\bf q}^\dag|0\rangle$ and the Goldstone
exciton $|1\rangle\!=\!S_-|0\rangle$ {\em represent at any nonzero ${\bf q}$, including the ${\bf q}\!\to\!0$ case, different
spin excitations}. Indeed, when calculating the action of the ${\hat {\bf S}}^2\!\equiv\!\frac{1}{2}(S_+S_-\!+S_-S_+)\!+{\hat
S}_z^2$ operator on the $|{\bf q};\!1\rangle$ state, then, by employing commutation equivalences $[S_+,{\cal
Q}_{\bf q}^\dag]\!\equiv\!{\cal A}_{\bf q}^\dag\!-{\cal B}_{\bf q}^\dag$ and $[S_-,{\cal Q}_{\bf q}^\dag]\!\equiv\!0$,
where the intra-sublevel operator is\vspace{-1mm}
\begin{equation}\label{A_operator}
{\cal A}_{\bf q}^{\dag}\!=\!\sum_{p}
e^{-iq_x\!p} a_{p+\frac{q_y}{2}}^{\dag}a_{p-\frac{q_y}{2}}\vspace{-2mm}
\end{equation}
(${\cal B}_{\bf q}^{\dag}$ means  $a\to b$ substitution), and, besides, by taking into account relations
$[S_-,{\cal A}_{\bf q}^\dag]\!\equiv\!{\cal Q}_{\bf q}^\dag$ and $[S_-,{\cal B}_{\bf q}^\dag]\!\equiv\!-{\cal Q}_{\bf
q}^\dag$, we obtain:\vspace{-2mm}
\begin{equation}\label{S_action}
{\hat {\bf S}}^2|{\bf q};\!1\rangle\!=\!\left[\left(\!\frac{{\cal N}_\phi}{2}\!\right)^2\!\!\!-\!\left(\!\frac{{\cal N}_\phi}{2}\!\right)\!+{\cal N}_\phi\delta_{{\bf q},0}\right]\!|{\bf q};\!1\rangle.\vspace{-2mm}
\end{equation}
It was also assumed that ${\cal A}_{\bf q}|0\rangle\!=\!{\cal N}_\phi\delta_{{\bf q},0}$ and ${\cal B}_{\bf
q}|0\rangle\!=\!0$ ($\delta_{{\bf q},0}\!\equiv\!\delta_{q_x\!,0}\delta_{q_y\!,0})$.  In all manipulations starting
from Eq. \eqref{QQ} we, certainly, took into account the `semi-classicality' of the Landau level, namely, the
inequalities ${\cal N}_\phi\!\gg\!1$ and $L\!\gg l_B$, by ignoring boundary effects. In particular,
semi-classicality means that the ${\bf q}\!\to\!0$ mathematical procedure, in common
units, implies $q\!\ll\!1/l_B$, whereas still $q>1/L$. So, one has to distinguish states ${\cal Q}^\dag_{{\bf q}\!\to
0}|0\rangle\!\equiv\!|{\bf 0};\!1\rangle$ and ${\cal Q}^\dag_{{\bf q}\equiv
0}|0\rangle\!\!\equiv\!S_-|0\rangle\!\equiv\!\!|1\rangle$, since the former, according to Eqs. \eqref{Sz_action} and
\eqref{S_action}, changes the spin numbers equally as compared to the ground state ($\delta S_z\!=\!\delta S\!=\!-1$),
while the latter changes only the $S_z$ component ($\delta S_z\!=\!-1$) and does not affect the $S$ number. The physical
meaning of the difference between the ${\bf q}\!\to\!0$ spin wave and the ${\bf q}\!\equiv\!0$ Goldstone exciton is
discussed in Appedix B. (See there also the comment on a similar property of an ordinary magnet described by the Heisenberg model.)

Now let us consider the state\vspace{-1mm}
\begin{equation}\label{n_q_state}
|{\bf q};\!n\rangle=(S_-)^{n-1}{\cal Q}_{\bf q}^\dag|0\rangle,\vspace{-2mm}
\end{equation}
where ${\bf q}\!\neq\!0$. As $S_-$ commutes with the ${\hat H}^{(1)}$ and ${\hat H}_{\rm Coul}$ operators, this state
is eigen, ${\hat H}_0|{\bf q};\!n\rangle=E_{n,{\bf q}}|{\bf q};\!n\rangle$ with energy\vspace{-1mm}
\begin{equation}\label{energy_n_q}
E_{n,{\bf q}}=\epsilon_{\rm Z}n+{\cal E}_{\bf q}+E_0.\vspace{-2mm}
\end{equation}
It is easy to calculate the corresponding spin quantum numbers and find $S_z\!=\!{\cal N}_\phi/2-n$, $S\!=\!{\cal
N}_\phi/2-1$.

Note that in the studied system the state with energy $E_{n,{\bf 0}}\!=\!\epsilon_{\rm Z}n+E_0$ is degenerate
as {\em two different} and even orthogonal {\em states $|{\bf 0};\!n\rangle$ and $|n\rangle$ have the same
energy}. The norm of state \eqref{n_q_state} is calculated with the help of Eq. \eqref{norm_n}, since that formula is
derived using the only property of $|0\rangle$:  the $S_z$ component in this state should be maximum, i.e.
$S_+|0\rangle\!\equiv\!0$. The $|{\bf q};\!1\rangle$ spin-wave state has the same property; therefore, writing $|{\bf
q};n\rangle=(S_-)^{n-1}|{\bf q};1\rangle$ and taking into account that \vspace{-1mm}
\begin{equation}\label{norm_sw}
\langle 1;{\bf q}|{\bf q};1\rangle\!\equiv\!{\cal N}_\phi,\vspace{-1mm}
\end{equation}
we find (with the substitutions $S\!\to\!{\cal N}_\phi/2-1$, $n\!\to\!n\!-\!1$ and $R_0\!\to\!{\cal N}_\phi$ in Eq.
\eqref{norm_n}) the squared norm:\vspace{-2mm}
\begin{equation}\label{norm_n_q}
{\widetilde R}_{{\cal N}_\phi,n}\!\!=\!\langle n;{\bf q}|{\bf q};n\rangle\!=\!\frac{{\cal N}_\phi!(n-1)!}{(\!{\cal N}_\phi\!-\!1)({\cal N}_\phi\!-n\!-\!1)!}\vspace{-1mm}
\end{equation}
which, note, is independent of ${\bf q}$.

The basis consisting of states $|n\rangle$ without spin-waves, and of $|{\bf q};n\rangle$ with a single spin wave is
formally incomplete in the context of a perturbation operator to be presented below.Owing to this, we should expand our
study by considering, for insyamce, double--spin-wave states ${\cal Q}_{\bf q'}^\dag{\cal Q}_{\bf
q}^\dag|n\!-\!2\rangle\equiv|{\bf q'}\!,{\bf q};n\rangle$ as well as states $|{\bf q''}\!,{\bf q'}\!,{\bf q};n\rangle$,
$|{\bf q'''}\!,{\bf q''}\!,{\bf q'}\!,{\bf q};n\rangle\,${\em etc\,}... Strictly speaking, these are not eigenstates of
the system due to the spin-wave exciton--exciton interaction. Such an interaction can be of two types: (i) a `kinematic
coupling' which takes place because exciton operators \eqref{QQ} obey an unusual commutation algebra (not belonging to
Bose or Fermi types, see Appendix A),  and (ii) a dynamic electro--dipole-dipole interaction since the spin-wave
exciton possesses dipole momentum $e l_B{\bf q}\times {\hat z}$ (it takes place for any magnetoexciton, see Refs.
\onlinecite{theory} and \onlinecite{go68}, and also discussion about the dynamic exciton-exciton scattering in
Ref. \onlinecite{di12}). In other words, the action of the Coulomb-interaction Hamiltonian on the $|{\bf
q}_2,{\bf q}_1;n\rangle$ state results not only in $\left({\cal E}_{{ q}_1}\!+{\cal E}_{{ q}_2}\!+E_0\!\right)|{\bf
q}_2,{\bf q}_1;n\rangle$ but also in an `additional' vector $[[{\hat H}_{\rm int},{\cal Q}_{{\bf q}_1}^{\dag}] {\cal
Q}_{{\bf q}_2}^{\dag}]|n\!-\!2\rangle$. The latter has a small norm -- by infinitesimal factor, $\lesssim E_{\rm
C}\!{\cal N}_\phi\!{}^{-1/2}$, different from the $|{\bf q}_2,{\bf q}_1;n\rangle$-state norm.

It is physically evident that the leading approximation in the framework of states with a minor number of spin-wave
excitons is, in fact, equivalent to the approximation of non-interacting spin excitons. This concerns both the dynamic
and kinematic interactions.\cite{foot2}  It is in in this `dilute regime' of non-interacting spin-wave excitons that we will
consider many-exciton states ${\cal Q}_{{\bf q}_1}^\dag{\cal Q}_{{\bf q}_2}^\dag...{\cal Q}_{{\bf q}_k}^\dag|n-k\rangle\!\equiv\!|\{{\bf q}_k\};n\rangle$ ($\{{\bf q}_k\}$ stands for a set of $k$ spin-wave excitons
with momenta ${\bf q}_{k},{\bf q}_{k\!-\!1},...{\bf q}_i,...{\bf q}_1$). These are definitely orthogonal: $\langle
m;\{{\bf q}'\!{}_{k'}\}|\{{\bf q}_k\};n\rangle\!\propto\delta\!{}_{\{{\bf q}'\!{}_{k'}\},\{{\bf q}_k\}}\delta_{m,n}$.
The quantum average of the $S_\perp$ transverse component vanishes if calculated in a `pure spin-wave' many-exciton state ${\cal Q}_{{\bf q}_1}^\dag{\cal Q}_{{\bf q}_2}^\dag...{\cal Q}_{{\bf q}_k}^\dag|0\rangle\!\equiv\!|\{{\bf
q}_k\};k\rangle$ (i.e. in the absence of Goldstone excitons). Moreover, for any arbitrary sets $\{{\bf q}_k\}$ and $\{{\bf q}'\!{}_{k'}\}$ we always
have\vspace{-1.5mm}
\begin{equation}\label{sw_zero1}
\langle k';\{{\bf q}'\!{}_{k'}\}|S_+|\{{\bf q}_k\};k\rangle\!\equiv\!0, \end{equation}
including the $\{{\bf q}'\!{}_{k'}\}\!\equiv\!\{{\bf q}_k\}$ case. By employing the
equations of Appendix A we can find the following matrix element in the dilute regime for states $|\{{\bf
q}_k\};n\rangle$ representing a `solution' of spin waves in `Goldstone-exciton condensate' ($n\!>\!k\!\ll\!{\cal N}_\phi$): \vspace{-1mm}
\begin{equation}\label{sw_zero2}
\begin{array}{r}
\!\!\!\langle n;\{{\bf q}'\!{}_{k'}\}|S_+|\{{\bf q}_k\};m\rangle\qquad\qquad\qquad\qquad\vspace{1.mm}\\{}\!\!\approx\langle k';\{{\bf q}'\!{}_{k'}\}|\{{\bf q}_k\};k\rangle\langle m|S_+|n\rangle\vspace{1.mm}\qquad\\\!\!\!\approx\langle k;\{{\bf q}{}_{k}\}|\{{\bf q}_k\};k\rangle\delta\!{}_{\{{\bf q}'\!{}_{k'}\!\},\{{\bf q}_k\!\}}\delta_{m,n\!+\!1}R_{{\cal N}_\phi,n\!+\!1}.
\end{array}\vspace{-0mm}
\end{equation}
Equations \eqref{sw_zero1} and \eqref{sw_zero2} mean that {\em the presence of spin waves is irrelevant to the
appearance of a $\langle S_\perp\rangle$ quantum average and effects related to azimuthal motion of the total spin}.

\vspace{-2mm}

\subsection{Perturbation term responsible for Goldstone mode stochastization}

\vspace{-2mm}

The key elementary process ensuring Goldstone mode stochastization is transformation of a Goldstone exciton into a
spin-wave and, thus, a change of the total spin number by $\delta S\!=\!-1$ at constant component $S_z$. The perturbation
field responsible for coupling between the $|n\rangle$ and $|{\bf q},n\rangle$ states should act on spin variables
(changing $S$) and violate translational invariance of the system resulting in appearance of excitations with nonzero
momenta ${\bf q}$. In this connection, spatial fluctuations of the effective Land\'{e} factor is just a relevant
perturbation, especially for GaAs/AlGaAs heterostructures. Indeed, in GaAs  the intrinsic spin-orbit
interaction of the crystal field with spins of conduction-band electrons changes significantly the effective $g$-factor
as compared to bare value $g_0\!=\!2$ resulting in a small total effective factor: $g\!\approx\!-0.43$ in bulk
GaAs. An external disorder field is added to the crystal one, therefore, small effective $g$ should in turn be
relatively well exposed to spatial disorder. Thus, considering $g=\langle g\rangle\!{}_{\mbox{\boldmath\tiny
$R$}}\!+g_1({\bf R})$, where the brackets $\langle...\rangle\!{}_{\mbox{\boldmath\tiny $R$}}\equiv\!\int\!...d{\bf R}/L^2$
mean spatial averaging, we get an additional perturbative term to Zeeman energy:\vspace{-2mm}
\begin{equation}\label{pert}
{\hat
V}_g\!=\frac{1}{2}\mu_BB\!\sum_i g_1({\bf R}_i)\sigma_{zi}.\vspace{-2mm}
\end{equation}
Therefore, the total Hamiltonian is\vspace{-1.5mm}
$$
  {\hat H}={\hat H}_0+{\hat V}_g.\vspace{-1.5mm}
$$

It is useful to employ Fourier expansion $g_1({\bf R})\!=\!\sum_{\bf q}\!e^{i{\bf qR}\!/l_B}\!\overline{g}_1({\bf q})$.
Let the $g$-disorder be spatially isotropic and, hence, characterized by correlator $G({R})\!=\!\int\!g_1({\bf
R}_0)g_1({\bf R}_0\!+\!{\bf R})d{\bf R}_0/L^2$; then the Fourier component $\overline{G}({ q})\!=\!\int\!G({
R})e\!{}^{-i{{\bf qR}/l_B}}d{\bf R}/(2\pi l_B)^2$ is also a function of the ${\bf q}$ modulus and $\overline{G}({
q})\!=\!{\cal N}_\phi|\overline{g}_1({\bf q})|^2/2\pi$. Following the common secondary quantization procedure,
${\hat{V}_g\!=\int\!{\hat \psi}^\dag({\bf R}){V}_g({\bf R}){\hat \psi}({\bf R})d{\bf R}}$, we obtain perturbation
in the form\vspace{-1mm}
\begin{equation}\label{V}
\hat{V}_g\!\!=\!\mu_BB\sqrt{\frac{\pi}{2{\cal N}_\phi}}\sum_{{\bf q}\neq 0}\!\Phi({ q})({\cal A}_{\bf q}^\dag\!-\!{\cal B}_{\bf q}^\dag).\vspace{-2mm}
\end{equation}
Here $\Phi({ q})=\sqrt{\overline{G}({ q})}\,{ L}_l(q^2\!/2)e^{-\!q^2\!/4}$, where ${ L}_l$ is the Laguerre polynomial. The
coupling is determined by matrix elements calculated with bra- and ket-vectors $|n\rangle$ and $|{\bf q};n\rangle$
where ${\bf q}\!\neq\!0$. We find $\langle n'\!;{\bf q}\,|\hat{V}_g|n\rangle\!\equiv\!\langle n;{\bf
q}\,|\hat{V}_g|n\rangle(1\!-\!\delta_{n,0})\delta_{n'\!,n}$, where \vspace{-1.5mm}
\begin{equation}\label{matrix}
\langle n;{\bf q}\,|\hat{V}_g|n\rangle=\!-\sqrt{\frac{2\pi}{{\cal N}_\phi}}\mu_BB\, n {\widetilde R}_{{\cal N}_\phi,n}\Phi({q})\vspace{-2mm}
\end{equation}
[see the squared norm \eqref{norm_n_q}]. Besides, we always have equivalence $\langle n'|\hat{V }_g|n\rangle\!\equiv\!0$.

\section{Spin-rotation (Goldstone) mode as an initial quantum state}

Macroscopically, the Goldstone mode is uniquely defined by total spin $S$ and angle $\theta$. However, quantum-mechanically the initial
$\theta$-deviation of the many-electron system may be organized in numerous ways.
Although the theoretical problem of studying the Goldstone-mode damping does not depend on the specific form of the
initial state, our first task is to microscopically model non-equilibrium $\theta$-deviation choosing
it with due account for the existing experimental results.\cite{Fukuoka,la15}

Considering combination $|C\rangle\!=\!\sum_nC_n|n\rangle$ and accounting for property ${\hat
S}_z|n\rangle\!=\!(S-n)|n\rangle$, one finds that, besides the normalization condition ${}\!\langle
C|C\rangle\!=\!1\!{}$, the set of coefficients ${C_n}$ must satisfy only one additional equation,\vspace{-1mm}
\begin{equation}\label{2.0}
\sum_nn|C_n|^2\langle n|n\rangle\!=\! S(1\!-\!\cos{\theta}),
\vspace{-3mm}
\end{equation}
in order to correspond to a Goldstone mode with parameters $S$ and $\theta$. It should also be remembered that any $|C\rangle$
vector is orbitally equivalent to the ground state. Indeed, if a 2D electron system is optically excited then
a certain ${\vec k}\!=\!0$ state can appear under condition\vspace{-1mm}
\begin{equation}\label{photon}
{\mathscr L}k_{{\rm photon}\parallel}\!\ll\!1,\vspace{-1mm}
\end{equation}
where length ${\mathscr L}$ is a characteristic of electron 2D-density spatial fluctuations  and $k_{{\rm
photon}\parallel}$ is the photon wave-vector component parallel to the electron system plane. (See discussion
concerning the value of ${\mathscr L}$ in Appendix B also referred to in Sec. V.) The ${\bold q}\!\equiv\!0$ state represented
by the $|C\rangle$ vector is an idealized model. In the known experimental research $\,$\cite{Fukuoka,la15} the emerging
spin-rotation state has a prehistory that consists of not only a very short stage of the immediate laser-pulse impact
changing the spin state, but also a longer stage of orbital relaxation preserving the total spin numbers. The
orbital relaxation, occurring during the time interval significantly shorter than spin stochastization and
resulting in the state which we consider as the initial one, includes `vertical' recombination
transitions$\,$\cite{foot0} and thermalization due to electron-electron and electron-phonon interactions. Ideally,
orbital relaxation should lead to the same orbital electron state that existed before the pumping laser pulse, i.e. to
the orbital state corresponding to the minimum of the total electrostatic energy. The latter {(determined by the
smooth random potential existing in the quantum well and by the {\em e-e} Coulomb correlations)} is the same as in the
$|0\rangle$ state.  We emphasize that { our state, described as the {initial} one in order to study spin
stochastization in the absence of any external influence, represents the {final} state of the preceding
orbital relaxation}. We do not know orbital relaxation details and, in principle, one cannot say {whether} after
such a relaxation prehistory the electron system comes exactly to a pure ${\bf q}\!\equiv\!0$ state. Our initial state
seems to be a combination of Goldstone $|n\rangle$ and spin-wave states $|\{{\bf q}_k\};n\rangle$.
However, {\em spin-waves are irrelevant to appearance of a transverse spin component and, therefore, to the observed
Kerr precession} [see Eqs. \eqref{sw_zero1} and \eqref{sw_zero2} and the related discussion in Sec. II]. So, for a
theoretical study of QHF spin-rotation dynamics it is quite relevant to consider only a $|C\rangle$ vector as an
initial state.

In a general case where the $|C\rangle$ state is not an eigen one for any ${\hat S}_{z'}$ operator, it should be called
an `unconventional' rotation-mode. Contrary to this, if, again, $|C\rangle$ represents microscopically the same
spin state but simply rotated as a `single-whole' state from the ${\hat z}$ direction to another direction ${\hat z}'$, then
it is natural to call this `rigid transformation' in the spin space (corresponding to global rotation of a `rigid'
ferromagnet) a `conventional' rotation-mode.\cite{con-spin-rot}
[The term `rotation-mode' accentuates the fact that every $|C\rangle$ state still remains diagonal for the ${\bf S}^2$ operator corresponding to its maximum value  ${\cal N}_\phi({\cal N}_\phi/2\!+\!1)/2$)]. The conventional and unconventional modes can be macroscopically characterized by the the same values of
${\bf S}$ and $\theta$ if only the $C_n$ coefficients satisfy equation \eqref{2.0}.

{To avoid misunderstanding, we note that in the conventional spin-rotation mode regardless of the laser-pulse intensity the spin-deviation angle $\theta$ is strictly equal to an angle $\beta$ given by the experimental setup. This fact strongly contradicts the considered experiments,\cite{Fukuoka,la15} where $\beta$ is the angle between ${\bf B}$ and the direction of the pumping laser beam. The measurements definitely show proportionality of the deviation $\theta$, and thereby of the precession amplitude, to the pulse intensity. That is, in these observations the angle $\theta$, being certainly much smaller than the given angle $\beta$, is strongly governed by intensity of the laser beam.}

\vspace{-6mm}

\subsection{One-photon absorption}

\vspace{-2mm}

Specific set $\{C_n\}$ must be additionally specified by microscopic initial conditions formulated appropriately to
the method of Goldston mode excitation. The laser pulse is formed with condensate of coherent photons equally
polarized and propagating at angle $\beta\!<\!90^\circ$ to the magnetic field, i.e., to the ${\bf S}$ direction in the
ground state. In real experimental geometry the laser beam is directed almost along the basic crystal axis
which, for its part, is perpendicular to the 2DEG plane. The total magnetic field ${\bf B}\!\parallel\!{\hat z}$ is
tilted by the angle $\beta$ from the normal to the 2DEG-plane. (The Landau-level functions and the
filling factor are determined by component $B_\perp\!=\!B\cos{\beta}$.) The laser pumping is in resonance with the
optical transition from the valence band to the electron Fermi edge corresponding at the $\nu\!=\!2l\!+\!1$ filling to
the spin-up (along the ${\bf B}$!) sublevel of the $l$-th Landau level. The absorbed photon with definite
angular momentum -1, i.e., antiparallel to the light-propagation direction,
results in appearance of a valence heavy-hole with total momentum $J=3/2$ and an electron with spin $S=1/2$, both in nonstationary states, oriented along the crystal axis (${\hat z}'$), inclined by $\beta$ to the ${\hat z}$ direction:$\,$\cite{Fukuoka,la15}
$J_{z'}=-3/2$ and $S_{z'}=1/2$. Thus, due to photon absorption
ensuring fast ($\sim 10\,$ps) electron-hole recombination processes,\cite{foot0} the spin state
$\uparrow=\!\left({1\atop 0}\right)$ of the spin-up electron in the conduction band is changed to the spin-rotated state
of the born electron$\,$\cite{LL3}\vspace{-0mm}
\begin{equation}\label{inclined}
{\nearrow}=\left(\cos{\!({\beta}\!/{2})} \atop  -{\sin{\!({\beta}\!/{2})}}\right)\vspace{-1mm}
\end{equation}
[$\beta$ is the Eulerian rotation angle; two others ($\alpha$ and $\gamma$) may be chosen equal
to zero]. The $\uparrow\,\to\!{\nearrow}$ replacement with the conservation of the orbital state of the total system
is a consequence of `verticality'  occurring owing to light absorption under the condition \eqref{photon}. The spin-up and spin-down probabilities for the `spin-inclined' state
{\footnotesize ${\nearrow}$}${}$ are $\cos^2\!{\!({\beta}\!/{2})}$ and $\sin^2\!{\!({\beta}\!/{2})}$, respectively. If
the electron system consists of $N_e>1$ spin-up electrons ($S_z\!=\!S\!=\!N_e/2$), then it is physically clear that, due
to absorption of one photon and subsequent `vertical' electronic processes, we get a $S_z$-non-diagonal
(`inclined') state with probability $\cos^2{\!(\!{\beta}\!/{2})}$ to have spin number $S_z\!=\!\!N_e\!/2$, with
probability $\sin^2{\!(\!{\beta}\!/{2})}$ to have $S_z\!=\!\!N_e\!/2\!-\!1$, and  any
$S_z\!<\!N_e\!/2\!-\!1$ values. At the same time, since the orbital state is not changed, such a `1-inclined' state should be a combination of a strictly spin-up state and a state arising due to a single action of the spin-lowering $S_-$ operator; therefore, it should remain diagonal for the ${\bf S}^2$ operator.

Let us consider $\nu\!=\!2l\!+\!1$ filling. The state with one `spin-inclined' electron if
simply written as
\begin{equation}\label{2.2}
|\!\uparrow\!\uparrow\!...\!\uparrow\!\!\mbox{\footnotesize${\nearrow}\!\!{}_{j}$}\!\uparrow\!...
\!\!\uparrow\rangle\equiv \left({\cos{\!\frac{\beta}{2}}-\!\sin{\!\frac{\beta}{2}}\,b^\dag_{p_{j}}\!a_{p_j}}\right)|0\rangle
\end{equation}
[$|0\rangle$ is ground state \eqref{2.1}]  is incorrect because it violates the principle of electron
indistinguishability and does not correspond to any definite value of conserved total spin $S\!=\!{\cal N}_\phi/2$.
However, every state \eqref{2.2} represents a correct combination in terms of the $S_z$ component:  the probabilities of
the $S_z\!=\!{\cal N}_\phi/2$ and $S_z\!=\!{\cal N}_\phi/2\!-\!1$ magnitudes are $\cos^2{\!(\!{\beta}/{2})}$  and
$\sin^2{\!(\!{\beta}/{2})}$, respectively. To describe correctly the `spin-inclined' state, adequate averaging of vectors \eqref{2.2} must be carried out where all individual spin-flip operators
$P_j^\dag\!\equiv\!b^\dag_{p\,{}_j}\!a_{p\,{}_j}$ participate equally. This collective state is obviously constructed
with the help of the $S_-\!=\!\sum_{j=1}^{{\cal N}_\phi}\!P_j^\dag$ operator, and, as a result, we obtain a correct
one-electron `spin-inclined' state\vspace{-2mm}
\begin{equation}\label{2.3}
|1,0\rangle=\left(\cos{\!\frac{\beta}{2}}-
\sin{\!\frac{\beta}{2}}\,{\cal N}_\phi^{-1/2}S_-\right)|0\rangle.\vspace{-1mm}
\end{equation}
Here it is taken into account that the squared norm of state $S_-|0\rangle$ is equal to ${\cal N}_\phi$.
Physically, Eq. \eqref{2.3} means that each of the ${\cal N}_\phi$ individual components $P_j^\dag|0\rangle$
contributes as the $1/{\cal N}_\phi$ part to collective one-electron spin-flip. In fact, the state described
by Eq. \eqref{2.3} and considered as an initial state of rotation-mode motion can be used in the case where the
number of `inclined' electron spins is much smaller than the number of electrons in the Landau level: $N\!\ll\!{\cal
N}_\phi$. Experimentally this situation is realized with a low-power laser pulse.\cite{la15}

\vspace{-3mm}

\subsection{Absorption of \mbox{\boldmath $N$} coherent photons}

\vspace{-2mm}

To describe the initial state in the $N\!\sim\!{\cal N}_\phi$ case, we generalize the above approach. First,
consider the opposite special case --- the situation with a maximum `quantum efficiency' of the laser pulse where $N\!=\!{\cal N}_\phi$
which means that all electron spins are aligned along ${\hat z}'$ tilted by angle $\theta\!=\!\beta$ to the ${\bf B}$
direction. A microscopic description of such a `conventional' rotation mode is$\,$\cite{con-spin-rot}\vspace{-2mm}
\begin{equation}\label{conventional}
|\mbox{\footnotesize${\nearrow}$}\!\mbox{\footnotesize${\nearrow}$}\!\!...\!\mbox{\footnotesize${\nearrow}$}
\rangle\!\equiv\!\displaystyle{\prod_{j=1}^{{\cal N}_\phi}\left({\cos{\!\frac{\beta}{2}}-\!\sin{\!\frac{\beta}{2}}\,P^\dag_{j}}\right)|0\rangle}.\vspace{-2mm}
\end{equation}
Going to the `unconventional' $N\!<\!{\cal N}_\phi$ case, first consider `conventional' rotation for subset
$\{J_N\}$ of $N$ electrons chosen among ${\cal N}_\phi$ ones:
$\left\{J_N\right\}\!\equiv\!\left\{p_{j_1}\!,\!p_{j_2}\!,...,\!p_{j_N}\right\}$ where the ordering
$p_{j_1}\!<\!p_{j_2}\!<...<\!p_{j_N}$ is assumed. Then such a `$J_N$-inclined' state (which is definitely not a
correct state describing the total system) is\vspace{-2mm}
\begin{equation}\label{2.4}
\begin{array}{l}
\displaystyle{\prod_{m=1}^{N}\left({\cos{\!\frac{\beta}{2}}-\!\sin{\!\frac{\beta}{2}}\,P^\dag_{j_m}}\right)|0\rangle
}\qquad\qquad\\\equiv\displaystyle{\sum_{n=0}^N\left(\cos{\!\frac{\beta}{2}}\right)\!\!{\vphantom{\left(\cos{\!\frac{\beta}{2}}\right)}}^{N-\,n}\!\!\left(\!-\sin{\!\frac{\beta}{2}}\right)^n
\frac{\left(S_{J_N\!-}\right)^n}{n!}}\,|0\rangle.
\end{array}\vspace{-1mm}
\end{equation}
Here $S_{J_N\!-}\!=\!\sum_{m=1}^{N}P^\dag_{j_m}$ is the spin lowering operator for the $\left\{J_N\right\}$ subset. The
norm of the state \eqref{2.4} is equal to one. It is noteworthy that each term of the right-hand-side of Eq. \eqref{2.4}
represents an eigenstate for the operator $S_z$ of the {\em total} system, namely:
$S_z(S_{J_N\!-})^n|0\rangle\!=\!(\!\frac{1}{2}{\cal N}_\phi\!-\!n)(S_{J_N\!-})^n|0\rangle$. [However,
$(S_{J_N\!-})^n|0\rangle$ is certainly not an eigenstate for total operator ${\bf S}^2$.] We note also that
expansion \eqref{2.4} over the $S_z$ eigenstates does not depend on specific subset $\left\{J_N\right\}$. Indeed, since
the squared norm of every $\left(S_{J_N\!-}\right)^n\!|0\rangle$ vector is  independent of $\left\{J_N\right\}$,
\begin{equation}\label{norm}
R_{Nn}\!=\!\langle 0|(S_{J_N\!+})^n(S_{J_N\!-})^n|0\rangle\!\equiv\!N!n!/(N\!-\!n)!
\end{equation}
[cf. Eq. \eqref{norm_n_q}], the norm of every item in the sum of Eq. \eqref{2.4} is
completely determined by numbers $N$ and $n$ only. In other words, the {\em quantum probability distribution over
the $S_z$ values given by Eq. \eqref{2.4} is determined only by number $N$} and does not depend on the choice of a specific
subset $\left\{J_N\right\}$. This probability,
\begin{equation}\label{prob}
\sin^{2n}{\!(\!{\beta}/{2})}[\cos^2{\!(\!{\beta}/{2})}]^{N\!-n}R_{Nn}\!/(n!)^2\,,
\end{equation}
 namely, the probability of the total $S_z$ component to take value ${\cal N}_\phi\!/2\!-\!n$, hence, it must also be
the same for the desired `$N$-inclined' state.

It is obvious that the generators for the $S_z$ eigenstates of the total system defined under the condition of the total $S$-number
conservation are the $(S_-)^n$ operators commuting with operator ${\bf S}^2$ [in contrast to operators
$(S_{J_N\!-})^n$ non-commuting with ${\bf S}^2$]. Now, in order to find the correct `$N$-inclined' state, we have to
take into account the {\em indistinguishability principle for various $\{J_N\}$ subsets} when $N$ coherent photons are
effectively absorbed allowing $N$ replacements ${\uparrow}\,\to\!{\nearrow}$. All possible samples $\{J_N\}$ must
equally contribute to the `$N$-inclined' state. We perform averaging over all the subsets by analogy with the above
transition from an individual spin-flip state $P_j|0\rangle$ to sum $\sum_jP_j|0\rangle\!\equiv\!S_-|0\rangle$ in
combination \eqref{2.3}. Now we consider transition from specific subset $\{J_N\}$ to sum over all possible
$\{J_N\}$. Note that there occurs equivalence\vspace{-2mm}
\begin{equation}\label{sum2}
\begin{array}{l}
\displaystyle{\!\!\sum_{\{J_N\}}\left(S_{J_N-}\right)^n
|0\rangle
\!\equiv\sum_{\{J_N\}}
\left(\sum_{m=1}^N\!P^\dag_{j_m}\right)\!\!\!{\vphantom{\left(\sum_{m=1}^NP^\dag_{j_m}\right)}}^n\!|0\rangle}
\vspace{0mm}\\
\!\displaystyle{=\!A(\!N,n\!)\!\left(\!\sum_{j=1}^{{\cal N}_\phi}P^\dag_{j}\!\!\right)\!\!\!{\vphantom{\left(\!\sum_{m=1}^N\!P^\dag_{j_m}\!\right)}}^n\!|0\rangle
\!=\!A(\!N,n\!)(S_-)^n|0\rangle}
\end{array}\vspace{-2mm}
\end{equation}
Therefore, averaging requires  replacement of the states $(S_{J_N-})^n |0\rangle$ in combination
\eqref{2.4} with the states $|n\rangle\!=\!(S_-)^n|0\rangle$. [Factor $A(\!N,n\!)$ can be calculated but is of no importance
for the following.] However, a simple $S_{J_N\!-}\!\to S_-$ substitution in Eq. \eqref{2.4} would certainly
be incorrect. The $\sim|n\rangle$ items in this combination should be appropriately normalized to satisfy the
condition above -- the probability for the $S_z$ component to be ${\cal N}_\phi\!/2\!-\!n$ must be determined by
value \eqref{prob}.  This condition provides an evident way to yield a proper collective `$N$-inclined' state: the
$\!(S_{J_N-})^n$-operators in the sum \eqref{2.4} must be replaced with the $({R_{Nn}/R_{N_\phi n}})^{1/2}(S_-)^n$
ones, where $R_{N_\phi n}\!=\!\langle n|n\rangle \!=\!\left.R_{Nn}\!\right|_{N\!=\!{\cal N}_\phi}$ That is, the
 correct `$N$-inclined' state representing the unconventional spin-rotation mode is\vspace{-4mm}
\begin{equation}\label{2.5}
|N,0\rangle=\sum_{n=0}^NC_n|n\rangle,\vspace{-3mm}
\end{equation}
where\vspace{-2mm}
\begin{equation}\label{2.6}
C_n\!=\!\frac{1}{n!}\!\left(\cos{\!\frac{\beta}{2}}\right)\!\!\!
{\vphantom{\left(\cos{\!\frac{\beta}{2}}\right)}}^{N\!\!-n}\!\!\!
\left(\!-\sin{\!\frac{\beta}{2}}\right)\!\!\!{\vphantom{\left(\cos{\!\frac{\beta}{2}}\right)}}^n
\!\sqrt{\frac{N!({\cal N}_\phi\!-\!n)!}{{\cal N}_\phi!(N\!-\!n)!}}.\vspace{-2mm}
\end{equation}
The $N\!=\!{\cal N}_\phi$ particular case corresponds to conventional mode \eqref{conventional}.

Concluding this section, it should be noted that equations \eqref{2.5}-\eqref{2.6} represent an expansion over the complete set
of orthogonal basis states -- the eigen states of the operator $S_z$ of the total electron system corresponding to the
${\bf q}\!\equiv\!0$ case and, besides, to fixed maximum value ${\bf S}^2\!=\!({\cal N}_\phi\!/2\!+\!1){\cal
N}_\phi\!/2$. The coefficients in this expansion are uniquely determined by the requirement to have a definite probability
distribution of the $S_z\!=\!{\cal N}_\phi\!/2\!-\!n$ eigenvalues stemming from the study of coherent spin-rotation
by the Eulerian angle $\beta$ of {\em any} $N$-electron subset ($0\!\leqslant\!N\!\leqslant\!{\cal N}_\phi$): the
probability is given by Eq. \eqref{prob} if $0\!\leqslant\!n\!\leqslant\!N$, or equal to zero if $n\!>\!N$. The
derivation of the state presented by Eqs. \eqref{2.5}-\eqref{2.6} is based on the assumption of
${\uparrow}\,\to\!{\nearrow}$ transition [see Eq. \eqref{inclined}] and on
the quantum-mechanical indistinguishability principle.

We will consider the Eq. \eqref{2.5} state with factors \eqref{2.6} as the initial one for the following
temporal state evolution. However, it will be shown that, in agreement with the macroscopic approach (Seq. I), the
relaxation law for transverse component $S_\perp$ is independent of specific set $\{C_n\}$.
\vspace{-3mm}

\section{Microscopic approach: precession without damping}

\vspace{-2mm}

Now we find non-stationary state $|N,t\rangle$ obeying  the Schr\"odinger equation $i\partial|N,t\rangle/\partial
t\!=\!{\hat H}_0|N,t\rangle$, where, as the first step, we consider only the
Hamiltonian \eqref{Hamiltonian} commuting with the $S_z$ and ${\bf S}^2$ operators. For the stationary $|n\rangle$ states we have: ${\hat
H}_0|n\rangle\!=\!(E_0\!+\!n\epsilon_{\rm Z})|n\rangle\equiv i\partial\!\left[ e^{-i(E_0\!+\!n\epsilon_{\rm
Z})t}|n\rangle\right]\!/\partial t$. As a result, if the initial state is
determined by Eq. \eqref{2.5}, the Schr\"odinger equation solution is\vspace{-2mm}
\begin{equation}\label{2.7}
|N,t\rangle=e^{-iE_0t}\sum_{n=0}^NC_n e^{-in\epsilon_{\rm Z} t}|n\rangle\vspace{-1.5mm}
\end{equation}

With the help of state \eqref{2.7}, we find quantum-mechanical averages of the relevant values at given instant $t$.
The total spin squared is a quantum number: ${\bf S}^2|N,t\rangle\!=\!\left[({\cal N}_\phi/2\!+\!1){\cal
N}_\phi/2\right]|N,t\rangle$ (i.e. $S\!=\!{\cal N}_\phi/2$). The average spin component $\langle S_z\rangle$ and the
average squares are also time-independent. For $C_n$ coefficients given by Eq. \eqref{2.6} we have: \vspace{-2mm}
$$
\langle t,N|S_z|N,t\rangle\!=\!{\cal N}_\phi/2-Nw,\,\;\; \langle
S_z^2\rangle\!=\!\langle S_z\rangle^2\!+w(1\!-\!w)N,\vspace{-2mm}
$$
and \vspace{-2mm}
$$\begin{array}{l}
\!\!\langle
S_x^2\!+\!S_y^2\rangle\!\equiv\!S(S\!+\!1)\!-\!\langle S_z^2\rangle\quad\vspace{1mm}\\
\qquad={\cal N}_\phi
Nw\!-\!(wN)^2\!-\!w(1\!-\!w)N\!+\!{\cal N}_\phi/2
\end{array}\vspace{-2mm}
$$
[where $w\!=\!\sin^2{\!(\!{\beta}/{2})}\!<\!1/2$]. In the framework
of the employed approximation neglecting any spin damping the only physical time-dependent value  is the
quantum average of transverse spin $\langle {\bf S}_\perp\rangle$. To obtain this, we calculate the
$\langle S_x\!+\!iS_y\rangle$ average:\vspace{-3mm}
 \begin{equation}\label{2.8}
 \begin{array}{l}
 \!\!\!\!\displaystyle{\langle t,N|S_+|N,t\rangle\!=\!e^{-i\epsilon_{\rm Z}t}\!\sum_{n=0}^{N\!-1}\!C_{n}^*C_{n\!+\!1}\langle n\!+\!1|n\!+\!1\rangle}\\\!\!\!\!\!\!\!\!\!\displaystyle{\quad=
 \!-\!\tan{\!(\!{\beta}/{2})}e^{-i\epsilon_{\rm Z}t}\,\sum_{n=0}^{N\!-1}\sqrt{\!(N_\phi\!-\!n)(N\!-\!n)}\,B_n},\vspace{-2mm}
 \end{array}
 \end{equation}
where\vspace{-2mm}
$$
B_n=\frac{N!}{n!\,(N\!-\!n)!}w^n(1\!-\!w)\!{}^{N\!-n}.\vspace{-2mm}
$$
So, with neglected damping Eq. \eqref{2.8} describes the Larmor precession in complete agreement with the equations of Sec. I . For the conventional Goldstone mode (i.e., for $N\!=\!{\cal N}_\phi$) the result is $|S_\perp|=({\cal
N}_\phi\sin{\!{\beta}})/2$ having an apparent geometric interpretation, see Fig. 1b. Considering the macroscopic limit
where $N\!\gg\!1$ while the $N/{\cal N}_\phi$ ratio is held constant, we notice that the $B_n$ numbers have a
sharp maximum at $n\!=\!Nw$ with width $\Delta n\sim \sqrt{N}$. Then summation over $n$ results in\vspace{-1mm}
\begin{equation}
\langle S_+\rangle\!\approx\!-\sin{\!(\!{\beta}/{2})}e^{-i\epsilon_{\rm Z}t}\sqrt{\! N({\cal N}_\phi\!-\!Nw)},\vspace{-1mm}
\end{equation}
and we make certain that macroscopically $|S_\perp|^2={\bf S}^2-S_z^2$; and the deviation angle is\vspace{-1mm}
\begin{equation}\label{theta}
\theta\!=\!\arccos{\!\left(1-\frac{2N}{{\cal N}_\phi}\sin\!{}^2{\frac{\beta}{2}}\,\right)}.\vspace{-1mm}
\end{equation}
For the conventional Goldstone mode we naturally get $\theta\!=\!\beta$.

It is interesting to consider the behavior of angle $\theta$ as a function of laser-pulse intensity, i.e. of the total
number of coherent photons ($I$) in the pulse. In case of weak intensity the number $N$ is simply
proportional to $I$. (This agrees with the experiments where the studied Kerr signal$\,$\cite{Fukuoka,la15} was found to be
proportional to the intensity of the laser beam.) Hence, if $N\!\ll {\cal N}_\phi$, we can write $dN/dI\!=\!W$, where
$W$ is a `quantum efficiency' factor independent of $N$. When speaking of `quantum efficiency' we consider not
the total number of absorbed photons but only a minute amount of them resulting in $\uparrow\,\to \nearrow$
 replacement in the conduction band (cf. Ref. \onlinecite{foot0}); so, of course, $W\!\ll\!1$. What happens with
growing intensity? It is clear that $N$ cannot exceed ${\cal N}_\phi$. In the case of $N$ comparable to ${\cal N}_\phi$
we have to take into account that the $\uparrow\,\to \nearrow$ replacement is realized only if the site in the Landau
level corresponding to relevant `vertical transition' $\,$\cite{foot0} is occupied by a spin-up electron
$\uparrow$. Indeed, the $\nearrow\,\to \nearrow$ process does not contribute to effective magnitude $W$ and,
therefore, the latter has to be proportional to the number of spin-up electrons, ${\cal
N}_\phi\!-\!N$, in the Landau level. Considering equation $dN/dI=W_0(1\!-\!N/{\cal N}_\phi)$ we find to within
unknown constant $W_0$ (which actually could be found experimentally by measuring $I$ and $\theta$) that
$N\!=\!{\cal N}_\phi[1\!-\!\exp{(-W_0I/{\cal N}_\phi)}]$. This equation, together with  Eq. \eqref{theta},
yields the $\theta(I)$ dependence.

\vspace{-1mm}

\section{Damping via stochastization due to smooth spatial disorder of $\mbox{\large{\boldmath $g$}}$-factor}

\vspace{-2mm}

In this
section we consider the problem in the `dilute regime', that is in the framework of the basis set where the characteristic number of spin-wave
excitons emerging due to stochastization is much smaller than the mean number of Goldstone excitons:
$k\ll\langle|S_\perp|\rangle\!\sim\! N$. Comparison of our approach at $k\!\ll\!{\cal N}_\phi$ with macroscopic
equation \eqref{1.2} enables to conclude that microscopically only the initial stochastization stage when
$t\!\ll\!T_2$ is studied, and therefore $T_2$ is determined by linear dependence
$|S_\perp(t)|=|S_\perp(0)|(1\!-\!t/T_2)$. To find this dependence (and thereby $T_2$), it is sufficient to study
$|n\rangle\!\to\!|{\bf q};n\rangle$ elementary transitions.

Thus, we now calculate the quantum mechanical average\vspace{-2mm}
\begin{equation}\label{avSperb}
\langle S_+\rangle={}_V\!\langle N,t|S_+|N,t\rangle\!{}_V,\vspace{-1mm}
\end{equation}
where state $|N,t\rangle\!{}_V$ obeys the equation \vspace{-0mm}
\begin{equation}\label{NSE}\vspace{-0mm}
i\partial|N,t\rangle\!{}_V/\partial t\!=\!({\hat H}_0\!+\hat{V}_g)|N,t\rangle\!{}_V
\end{equation}
[see \eqref{Hamiltonian} and \eqref{V}] that should be solved by projecting onto the Hilbert space determined by orthogonal basis vectors $|n\rangle$ and
$|{\bf q},n\rangle$. The initial condition is given by equation $|N,0\rangle\!{}_V\!=\!|N,0\rangle$ [see Eqs.
\eqref{2.5}-\eqref{2.6}]. Then searching for the solution in the form\vspace{-2mm}
\begin{equation}\label{sol}
\begin{array}{r}
\displaystyle{|N,t\rangle\!{}_V=e^{-iE_0t}\sum_{n=0}^NC_n e^{-in\epsilon_{\rm Z} t}\left[\vphantom{\sum_{\bf q}e^{-i{\cal E}_{ q}t}b_{n{\bf q}}(t)|{\bf q},n\rangle}a_n(t)|n\rangle\right.}\quad\\
+\displaystyle{\left.\sum_{\bf q}e\!{}^{-i{\cal E}_{ q}t}b_{n{\bf q}}(t)|{\bf q},n\rangle\right]},
\end{array}\vspace{-2mm}
\end{equation}
where $a_n(0)\!=\!1$, and $b_{n{\bf q}}(0)\!=\!0$, and substituting this into Eq. \eqref{NSE}, we come, with the help of
Eqs. \eqref{V} and \eqref{matrix}, to
\vspace{-1mm}
\begin{equation}\label{a_eq} {}\!\!i\partial a_n/\partial
    t=\!\displaystyle{\langle n|n\rangle}^{-1}\!\sum_{{\bf q}}{e\!{}^{-i{\cal E}_{ q}t}{\langle n|\hat{V}_g|{\bf
    q},n\rangle}b_{n{\bf q}}(t)}\vspace{-5mm}
\end{equation}
and
\vspace{-2mm}\begin{equation}\label{b_eq}
i\partial b_{n{\bf q}}/\partial t\!=\!\langle n;{\bf q}\,|{\bf q};n\rangle\!{}^{-1}e^{i{\cal E}_{ q}t}{\langle n;{\bf q}|\hat{V}_g|n\rangle}a_n(t)\,.\vspace{-1mm}
\end{equation}
The studied initial stage, $t\ll T_2$, actually means condition $|b_n|\!\ll\!|a_n|$ in this case, i.e. we have to find
the solution of Eqs. \eqref{a_eq} in the leading approximation in perturbation ${\hat V}_g$. To be more precise, $b_{n{\bf
q}}$ must be calculated to the first order and $a_n$ to the second-order (both corrections are essential since the
contribution to stochastization is determined by the terms in $a_{n\!+\!1}^*a_n$ and $b_{n\!+\!1{\bf q}}^*b_{n{\bf
q}}$ proportional to $V_g^2$). So,
\vspace{-3mm}
\begin{equation}\label{b_n}\vspace{-3mm} b_{n{\bf q}}\!=\!nf_b({\bf
q},t)\;\;\mbox{and}\;\; a_n\!=1\!+\!n\!\left(\!1\!-\!\frac{n}{N_\phi}\right)\!f_a(t),
\end{equation}
where\vspace{-3mm}
$$
\begin{array}{l}
\!\!\displaystyle{f_b({\bf q},t)\!=\sqrt{\frac{2\pi}{{\cal N}_\phi}}\frac{\mu_BB}{{\cal E}_{ q}}\left(e{}^{i{\cal E}_{ q}t}\!\!-1\right)\!\Phi({ q})}\quad\mbox{and}\quad\vspace{2mm}\\
\!\!\displaystyle{f_a(t)\!=\!\frac{2\pi(\mu_BB)^2}{{\cal N}_\phi}\!\sum_{\bf q}|\Phi({ q})|^2\!\!\int_0^t\!\!i\frac{1\!-e\!{}^{-i{\cal E}_{ q}t'}}{{\cal E}_{ q}}\,dt'.}
\end{array}\vspace{-2mm}
$$
Substitution of Eq. \eqref{b_n} into Eq. \eqref{sol} and then into Eq. \eqref{avSperb} yields\vspace{-2mm}
$$
\langle S\!{}_+\!\rangle
\!=\!e^{-i\epsilon_{\rm Z}t}\!\displaystyle{\sum_{n=0}^{N\!-1}\!\!C_n^*C_{n\!+\!1}\langle n\!\!+\!1|n\!\!+\!1\rangle\!\!\left(\!\!1\!+\!f_a\!\!-i\frac{2n}{{\cal N}_\phi}{\rm Im}f_a\!\!\right)}\!.\vspace{-2mm}
$$
The imaginary part of $f_a$ results only in an inessential correction to the frequency of Larmor oscillations
$\epsilon_{\rm Z}$ and does not contribute to damping. By ignoring ${\rm Im}f_a$ the expression in the parentheses
ceases to depend on $n$. Then we find $\langle S\!{}_+\!(t)\rangle$ proportional to $S\!{}_\perp(0)=\langle
S\!{}_+\!(0)\rangle$ that means that the transverse relaxation process occurs in the same way regardless of the specific value
of initial deviation. This result is certainly in agreement with the macroscopic approach results. So, we
obtain \vspace{-1mm}
\begin{equation}\label{Splus}
\langle S\!{}_+\!(t)\rangle\!=S\!{}_\perp(0)e^{-i\epsilon_{\rm Z}t}\left[1\!+\!{\rm Re}f_a(t)\right],\vspace{-2mm}
\end{equation}
where\vspace{-1mm}
\begin{equation}\label{Ref_a}
\begin{array}{l}
\!\!\!\!{\rm Re}f_a(t)\\\vspace{.5mm}\!\!\!\!\!\vspace{-1mm}=\!-2\pi(\mu_BB)^2\!\!\displaystyle{\int_0^{{\cal E}\!\!{}_\infty}}
\!\!\!\!\!\!\!\Phi^2(q)\!\left[1\!-\!\cos{\!({\cal E}_qt)}\right]\!{\nu({\cal E}_q\!)d{\cal E}_q}\!/{{\cal E}_q}\!\!{}^2\vspace{-1mm}
\end{array}
\end{equation}
[$\nu(\varepsilon)$ denotes the density of states: $\nu({\cal E}_q)\!=\!qdq/d{\cal E}_q$, in particular
$\nu(0)\!=\!M_{\rm x}$]. Generally, any further transformation of expression \eqref{Ref_a} requires a more
detailed description of the $\Phi(q)$ and $\nu(\varepsilon)$ functions which in turn are determined by the $g$-factor
spatial disorder and by the real size-quantized (along the perpendicular ${\hat z}$ direction) electron wave-function
in the quantum well. However, at sufficiently large times $t$, when condition ${\cal E}_qt\gtrsim
1$ means that $q\ll 1$ and $\Phi(q)\approx \Phi(0)$, then ${\rm Re}f_a(t)=-[\pi\mu_BB\Phi(0)]^2M_xt$. If one recalls
the definition of $\Phi(q)$ via  $G$-correlator, then simple analysis shows that this asymptotic expression is valid if
$t\gg M_{\rm x}(\Lambda/l_B)^2$, where $\Lambda$ is the characteristic correlation length of smooth spatial
disorder. Thus, performing comparison with Eq. \eqref{Splus}, we find the formula
$|S\!{}_\perp\!(t)|=|S\!{}_\perp\!(0)|(1-t/T_2)$ with inverse stochastization time\vspace{-1mm}
\begin{equation}\label{T_2}
1/T_2=[\pi\mu_BB\Phi(0)]^2M_{\rm x}.\vspace{-1mm}
\end{equation}
 This result is valid within the time interval $M_{\rm x}(\Lambda/l_B)^2\ll t\ll T_2$.

As examples, we study two specific kinds of random spatial function $g_1({\bf r})$ distribution. For simplicity, we consider
the most `strong ferromagnet' state of unit filling where the spin-up sublevel of the zero Landau level is
completely occupied and other electron quantum states are empty, i.e. $l\!=\!0$.

\vspace{-2mm}

 \subsection{Gaussian disorder}

 \vspace{-1mm}

First, let the correlator be Gaussian, $G({\bf r})\!=\!\Delta_g^2e^{-r^2/\Lambda^2}$ being parameterized by fluctuation amplitude
$\Delta_g$ and correlation length $\Lambda$. Then $\Phi^2(q)\!=\!\displaystyle{{\Delta_g^2\Lambda^2}e^{-\lambda
q^2}\!\!\!/{4\pi l_B^2}}$, where $\lambda\!=\!1/2\!+\!\Lambda^2\!/4l_B^2$. In accordance with the actual situation, one may consider
$\lambda\!\gg\!1$; in this case the characteristic values are $q\!\ll\!1$ and we may again put $\nu({\cal
E}_q)\!\approx\!M_{\rm x}$ and integrate in Eq. \eqref{Ref_a} from 0 to $\infty$. Then we obtain\vspace{-1mm}
 $$
 \begin{array}{l}
 \!\!\!{\rm Re}f_a=-\left(t/T_2^{(G)}\right)\displaystyle{\left\{(2/\pi)\arctan{\left(2t/\tau_0^{(G)}\right)}
 \vphantom{\ln{\left[1\!+\!\left(t/{\tau_0^{(G)}}\right)^2\right]}}\right.}
 \vspace{.5mm}
 \\{}\qquad\quad-
 \left.\left(\tau_0^{(G)}\!\!/2\pi t\right)\ln{\!\left[1\!+\!\left(2t/{\tau_0^{(G)}}\!\right)^2\right]}
 \right\},
 \end{array}\vspace{-4mm}
 $$
 where\vspace{-1mm}
 \begin{equation}\label{T_2gauss}
 1/T_2^{(G)}=\pi M_{\rm x}(\mu_BB\Lambda \Delta_g/2l_B)^2\vspace{-0mm}
 \end{equation}
(due to the misprint in Ref. \onlinecite{la15}, this expression is by a factor of $1/2$ different from the result given there); and at the
initial stage the dependence is quadratic ${\rm Re}f_a\!\approx\! -2t^2\left[\pi T_2^{(G)}\tau_0^{(G)}\right]^{-1}$
where the characteristic transient-stage time is\vspace{-1mm}
\begin{equation}\label{taugauss}
\tau_0^{(G)}=M_{\rm x}(\Lambda\!/l_B)^2.\vspace{-2mm}
\end{equation}

\vspace{-2mm}

\subsection{Lorentzian disorder}

 \vspace{-1mm}

If the correlator is determined by the Lorentz distribution, $G({ r})\!=\!\Delta_g^2(r^2\!/\Lambda^2\!+\!1)^{-1}$, then
$\Phi^2\!(q)\!=\!(\Delta_g\Lambda\!/l_B)^2K_0(q\Lambda\!/l_B)e^{-q^2\!/2}\!\!/2\pi$ ($K_0$ is the Bessel function), and
$\left.\Phi(q)\right|_{q\to 0}$ in Eq. \eqref{T_2} logarithmically goes to infinity. In this case it is necessary to take into
account a real minimum of $q$'s which is determined by uncertainty $\delta q\!\sim\!M_{\rm x}l_B|\nabla\varphi|$
related to violation of the translational invariance owing to smooth random electrostatic potential $\varphi({\bf r})$
inevitably existing within the 2D channel (see, e.g., Ref \onlinecite{zh14} and references therein; indeed,
$|\nabla\varphi|\Lambda\sim\!0.4-0.6\,$meV). So, substituting $q_{\rm min}\!\sim \delta q$ instead of zero in Eq.
\eqref{T_2}, for $T_2$ in the case of the Lorentz disorder\vspace{-2mm} we find
\begin{equation}\label{T_2lorentz}
{1}/{T_2^{(L)}}\!=\!\frac{\pi}{2}(\mu_BB\Delta_g\Lambda\!/l_B)^2M_{\rm x}\ln{\!\frac{2}{M_{\rm x}\overline{\varphi}}},\vspace{-2mm}
\end{equation}
where $\overline{\varphi}$ is the smooth random potential amplitude,
$\overline{\varphi}\!=\!\langle|\nabla\varphi|\Lambda\rangle$. As again $\Lambda\!\gg l_B$, for the initial stage of
stochastization one can calculate the integral in Eq. \eqref{Ref_a} by putting $\nu({\cal E}_q)\!=\!M_{\rm x}$ and
${\cal E}_\infty\!=\!\infty$, and at $t\lesssim \tau_0^{(L)}\!$ find that ${\rm Re}f_a\!\approx\!
-t^2\!/T_2^{(L)}\!\tau_0^{(L)}$, where\vspace{-2mm}
\begin{equation}\label{tauL}
\tau_0^{(L)}=\sqrt{2}M_{\rm x}(\Lambda\!/l_B)^2\ln{\!\frac{2}{M_{\rm x}\overline{\varphi}}}\,.\vspace{-2mm}
\end{equation}

\section{Kinetic approach to the stochastization problem}

\vspace{-2mm}

In the previous sections the purely quantum-mechanical problem of excitation evolution has been solved. When so doing
only the initial stage is relevant and has been considered. Except for a short interval of
the transition process, this stage of transverse relaxation is described by a linear function of time. Generally we
have no reasons to think that the dependence $|S_\perp(t)|$ becomes а damping exponent for longer times $t\!\gtrsim\!T_2$ --
as it would follow, for instance, from phenomenological equation \eqref{1.2}. As mentioned above, a complete solution of the
quantum-mechanical problem requires consideration of states\vspace{-1mm}
\begin{equation}\label{state-q}
|\{{\bf q}\}_k;n\rangle=
\left(S_-\right)^{n\!-\!k}\!{\cal Q}_{{\bf q}_1}^\dag\!{\cal Q}_{{\bf q}_2}^\dag...{\cal Q}_{{\bf q}_k}^\dag|0\rangle\vspace{-1mm}
\end{equation}
(see the II-B subsection). In the presence
of perturbation responsible for $|\{{\bf q}\}_k;n\rangle\to |\{{\bf q}\}_{k\!+\!1};n\rangle$ transitions
occurring within the `$n$-shell' (i.e. at a constant total number of excitons $n$) an effective number of spin-wave
excitons $k$ grows in time, and in the case $k\!\sim\!n\!\sim\!N\!\sim {\cal N}_\phi$ our model of non-interacting spin
excitons fails.\cite{foot2} Then, certainly, the stochastization process {\em a priori} becomes non-exponential.

If $k\ll {\cal N}_\phi$ (which is definitely valid for small deviations at the initial time, i.e. if $N\ll {\cal
N}_\phi$), then the state \eqref{state-q} is quite meaningful and represents spin-wave exciton gas in the `dilute
limit'. In this section we demonstrate a kinetic approach to the stochastization problem and consider state
$|n\rangle$ as the initial one with number $n\gg 1$ in the vicinity of the maximum:
$n\!\approx\!n_m\!=\!N\sin\!{}^2{(\!{\beta}\!/{2})}$, and still consider $n\!\ll\!{\cal N}_\phi$. Following the decay
mechanism related to transitions $|n\rangle\!\to\!|{\bf q}_1;n\rangle\!\to\!|\{{\bf q}\}_2;n\rangle...$, we study the
$|\{{\bf q}\}_k;n\rangle\!\to \!|\{{\bf q}\}_{k\!+\!1};n\rangle$ process and the corresponding change of value
$S_\perp^2\!=\!(S_+S_-\!+\!S_-S_+)/2$. The operator $S_\perp^2$, if considered within the `dilute limit', is diagonal in
the basis consisting of states \eqref{state-q}. Taking into account formula\vspace{-2mm}
\begin{equation}\label{normk}
\begin{array}{l}
\displaystyle{\langle n;\{{\bf q}\}_k|\{{\bf q}\}_k;n\rangle}\qquad\qquad\qquad\vspace{2mm}\\
\displaystyle{\approx\frac{(n\!-\!k)!({\cal N}_\phi\!-\!2k)!}{(\!{\cal N}_\phi\!\!-\!n\!\!-\!k)!}{\langle k;\{{\bf q}\}_k|\{{\bf q}\}_k;k\rangle}}
\end{array} \vspace{-2mm}
\end{equation}
[see Eq. (A.6) in Appendix A and cf. Eq. \eqref{norm_n_q}], we obtain the semi-classical value
\begin{equation}\label{Sperp}
S_\perp^2\!\!=\!\frac{\langle n;\!\!\{{\bf q}\}_k|S_+S_-\!\!\!-\!S_z|\{{\bf q}\}_k;\!n\rangle}
{\langle n;\{{\bf q}\}_k|\{{\bf q}\}_k;n\rangle}\!\!
\approx\!{\cal N}_\phi\!\left(n\!-\!k\right).
\end{equation}
This formula reveals that the transverse spin-component squared is proportional
to $n\!-\!k$ which is the number of Goldstone spin excitons. Its decrease  (the increase in $k$) determines the transverse relaxation process.

Now let us find the rate of the $S_\perp^2$ change by calculating total probability for transformation of the
$|i\rangle\!=\!|\{{\bf q}\}_k;n\rangle$ state into various states $|f_{\bf q}\rangle\!=\!|\{{\bf q}\}_{k+\!1};n\rangle$
per unit time (considering ${\bf q}_{k+\!1}\!=\!{\bf q}$). This probability is equal to the growth rate of number
$k$, \vspace{-3mm}
\begin{equation}\label{dk/dt}
dk/dt= \sum_{\bf q}W_{i\to f_{\bf q}},\vspace{-3mm}
\end{equation}
where partial probabilities are determined by the well known formula\vspace{-2mm}
\begin{equation}\label{W}
W_{i\to f_{\bf q}}=\!\frac{2\pi|\langle f_{\bf q}|{\hat V}_g|i\rangle|^2}{
\langle i|i\rangle\langle f_{\bf q}|f_{\bf q}\rangle}\,\delta(E_{f_{\bf q}}-E_i),
\vspace{-2mm}
\end{equation}
where we again use operator \eqref{V} as a perturbation. In the framework of our approximation,
$k\!<\!n\!\ll\!{\cal N}_\phi$, the matrix element is \vspace{-2mm}
\begin{equation}\label{element}
\langle f_{\bf q}|{\hat V}_g|i\rangle\!\approx\!-2(n\!-\!k)\mu_B\!B
\sqrt{\frac{\pi}{2{\cal N}_\phi}}\,\Phi({ q})\langle f_{\bf q}|f_{\bf q}\rangle.
\vspace{-2mm}
\end{equation}

The sum in Eq. \eqref{dk/dt} represents summation over nonzero ${\bf q}$'s. It looks, however, rather
uncertain since formally the $\delta$-function argument in Eq. \eqref{W} is equal to $q^2\!/2M_{\rm x}$. A more detailed study enables
us to eliminate this uncertainty (see Appendix C) and finally obtain, with the help of Eqs. \eqref{Sperp} --
\eqref{element} and Eq. (A.4), the kinetic equation describing the damping process:\vspace{-3mm}
\begin{equation}\label{kin_eq}
dS_\perp^2\!/dt=-2S_\perp^2\!/T_2.\vspace{-2mm}
\end{equation}
The derived equation is independent of $n$ and $k$, and the transverse relaxation time $T_2$ is just the same as that
given by equation \eqref{T_2} in Section V, including particular cases \eqref{T_2gauss} and \eqref{T_2lorentz}.
So, if the initial deviation from the equilibrium direction is small, $|S_\perp(0)|\ll {\cal N}_\phi$, then the kinetic
equation \eqref{dk/dt} results in exponential damping of the Kerr rotation:\vspace{-3mm}
\begin{equation}\label{exp_damping}
 |S_\perp(t)|=|S_\perp(0)|e^{-t/T_2}.\vspace{-2mm}
\end{equation}
(The transient stage occurring in time $t\lesssim \tau_0$ is certainly not described in the framework of the kinetic approach.)

\vspace{-3mm}

\section{Conclusion}

\vspace{-2mm}

The study addresses a spin-rotation mode emerging at optical excitation in quantum Hall
spin-polarized systems. This mode is macroscopically indistinguishable from a simple turn of the entire electron spin
system from the ${\hat z}$-direction. However, the general phenomenological approach shows that the damping of the spin-rotation precession in the quantum Hall ferromagnet hardly obeys the Landau-Lifshitz equation. The microscopic approach reveals that
the quantum state of the unconventional spin-rotation mode is not equivalent to rotation as a single-whole of all spins by the
same angle. This specific property manifests itself in the dependence of the effective (macroscopic) rotation angle $\theta$ on
laser pumping intensity rather than on the laser-beam direction alone. [See Eq. \eqref{theta} where the number $N$ is
determined by laser pumping; if $N$ reaches ${\cal N}_\phi$, then the unconventional mode becomes a conventional
Goldstone mode and the equality $\theta\!=\!\beta$ holds even at higher intensities of laser pumping.]

One can note a similarity between the optically-induced spin-dynamics in a QHF and in dielectric magnets where spin precession
occurs also owing to `coherent magnon generation'.\cite{femtosecond} {Indeed, such generation resulting in coherent spin precession appears due to an `optomagnetic interaction' with media if the pumping laser beam is inclined at an angle with respect to the magnetization axis (cf. our angle $\beta$); i.e. the experimental technique is similar to the precession excitation in Refs. \onlinecite{Fukuoka} and \onlinecite{la15}. Besides, it is possible to assume that the unconventional spin-rotation mode would be an adequate microscopic description [see Eqs. \eqref{spin-rot}, \eqref{n-states} and \eqref{2.7}] for the coherent precession state studied in dielectric magnets in works \onlinecite{femtosecond}. The initial rotation angle in those experiments is proportional to laser intensity as in the works$\,$\cite{Fukuoka,la15} with a QHF.} Without going into a discussion on the optomagnetic interaction,\cite{femtosecond} we notice that our situation with appearance of the QHF spin-rotation mode looks still more transparent since it is based on a purely electronic pattern. In our case it is the reaction of a strongly correlated electron gas described in terms of collective eigenstates to an elementary single-electron process representing simple replacement of a spin-polarized conduction-band electron with orbitally the same but `spin-inclined' one generated by an absorbed photon (see Refs. \onlinecite{la15} and \onlinecite{foot0}). {Finally, note the following: two types of magnons, -- the Godstone one (with ${\bf q}\!\equiv\!0$, $|\delta S_z|\!=\!1$ and $\delta S\!=\!0$) and the spin wave (with ${\bf q}\!\neq\!0$ and $|\delta S_z|\!=\!\delta S\!=\!1$), -- do exist also in common dielectric magnetics described by the Heisenberg Hamiltonian (see Appendix B).}

Our microscopic approach consists in solving a non-stationary Schr\"odinger equation where the
unconventional spin-rotation mode is considered as the initial state.   As a perturbation resulting in damping,
the stochastization mechanism {is studied which is related} to spatial fluctuations of the effective Land\'e factor.
Those are most likely related to spatial fluctuations of 2DEG thickness, since the effective $g$-factor of 2D electrons
depends on the quantum well width.\cite{iv} Meanwhile the spatial fluctuations of the width also affect 2D electrons as an
additional effective electric field contributing thereby to the effective smooth random potential.
Thus, the correlation length of the $g$-factor fluctuations $\Lambda$ is supposed to be approximately equal to the correlation length
of the smooth random potential in the quantum well, $\sim\!50\,$nm. Assuming $g$-fluctuation amplitude $\Delta_g\!\sim\!\!0.005$, which seems fairly realistic,\cite{iv} we find characteristic damping time $T_2\sim 1-10\,$ns according to Eqs.  \eqref{T_2gauss} and
\eqref{T_2lorentz}. (We also used $1/M_{\rm x}\!\sim\! 2\,$meV
and $B\!=\!3-10\,$T in accordance with the available experimental data.\cite{ga08,la15}) This transverse spin
relaxation time is much shorter than total relaxation time in  a similar system.\cite{zh14} The microscopic
approach also enables us to describe the transient process preceding establishment of the linear time dependence of
diminishing transverse component $|S_\perp|$. The characteristic time of this short transient stage is given by
Eqs. \eqref{taugauss} and \eqref{tauL}, being $\tau_0\!\sim\!10-50\,$ps.

The kinetic approach also shows that, as expected, at small initial excitations the damping process for times
$t\!\gg\!\tau_0$ occurs exponentially [see Eq. \eqref{exp_damping}] just with the $T_2$ time calculated in the
framework of the solution of the non-stationary Schr\"odinger equation.

Formally, the results reported are applicable only in a narrow region near integer fillings 1 and 3 (although, in
principle, they seem to be phenomenologically projected onto the case of fractional ferromagnets where $\nu=1/3,1/5,...$;
cf. research in Ref. \onlinecite{di12,di11}). Meanwhile it is known that a skyrmion texture with well reduced
spin-polarisation emerges even at a small deviation of the filling factor from 1. Theoretically this `skyrmionic'
ferromagnet becomes `softer' than the unit-filling one, and the Goldstone mode damping should occur much faster due to
appearance of additional stochastization channels related to some soft modes forbidden in the
integer-filling state. This theoretical view is confirmed experimentally by both the observation of Goldstone mode
dynamics$\,$\cite{la15} and by the study of total spin relaxation (recovery of the ${\bf S}$ vector to the ground
state magnitude) in a quantum Hall ferromagnet.\cite{zh14}

In conclusion, we note that the work presented is done by taking into account the experimental background dealing with
`classical' quantum Hall systems, i.e., created in GaAs/AlGaAs structures. Nevertheless, our approach and the results
obtained could be actual or/and at least useful as a basis for future studies of more up-to-date quantum-Hall-ferromagnet
states (in graphene, in ZnO/MgZnO structures, etc.), which have been lately studied intensively, yet, in the absence of
relevant data on relaxation of collective spin states.

The research was supported by the Russian Foundation for Basic Research: Grant 18-02-01064.

\vspace{-2mm}
\appendix

\vspace{-2mm}
\section{}

\vspace{-2mm}

In the equations presented in this section we do not make any formal difference between Goldstone and spin-wave excitons, that is ${\bf q}$
may be exactly equal to zero: $S_-\!\equiv\!{\cal Q}_{0}^\dag$. In the QHF ground state both have equal norms: $\langle
S_+S_-\rangle\!=\!\langle {\cal Q}_{\bf q}{\cal Q}_{\bf q}^\dag\rangle\!\equiv\!{\cal N}_\phi$ (here and everywhere
below $\langle...\rangle$ means averaging over the ground state: $\langle...\rangle\!\equiv\!\langle 0|...|0\rangle$).

First, we write out the QHF Hamiltonian [see \eqref{Hamiltonian} and \eqref{operators}] in terms of the so-called `excitonic representation' within the two-sublevel approximation relevant to calculate the spin-wave exciton energy to first order in Coulomb coupling. Omitting all the terms commuting with the ${\cal Q}_{\bf q}^\dag$ operator [in particular, we also omit the ${\hat H}\!{}^{(1)}\!\!=\!\omega_c({\cal A}_0\!+\!{\cal B}_0)\!/2$ term], we get a reduced secondary-quantization form of the Hamiltonian:\vspace{-2mm}
$$
\begin{array}{l}
\displaystyle{{\hat{\cal H}}_0'=-\epsilon_{\rm Z}{\hat S}_z\!+\!{\hat H}_{\rm Coul}'=-\epsilon_{\rm Z}(\!{\cal A}_0\!-\!{\cal B}_0)\!/2\qquad\qquad}\vspace{2mm}\\\qquad\displaystyle{+ \frac{1}{2}\!\sum_{\bf q}\!{\widetilde U}(q)({\cal A}_{\bf q}^\dag{\cal A}_{\bf q}+2{\cal A}_{\bf q}^\dag{\cal B}_{\bf q}+{\cal B}_{\bf q}^\dag{\cal B}_{\bf q})},
\end{array}\vspace{-3mm}
$$
where $\displaystyle{{\widetilde U}(q)\!=\!e^{-q^2\!/2}[{ L}_l(q^2\!/2)]^2\!\int\!U(r)e^{{\rm i}{\bf qr}}d{\bf r}\!/2\pi}$ (${ L}_l$ is the Laguerre polinomial). Then using the commutation rules\vspace{-1mm}
$$
e^{i\phi}\!\!\left[{\cal A}_{\bf q_1}^\dag,
  {\cal Q}_{{\bf q_2}}^\dag\right]\!=\!
  -e^{-i\phi}\left[{\cal B}_{\bf q_1}^\dag,
  {\cal Q}_{{\bf q_2}}^\dag\right]\!=\!
  -{\cal Q}_{{\bf q_2\!+\!q_1}}^\dag \eqno({\rm A}.1) \vspace{-2mm}
$$
and\vspace{-2mm}
$$
\displaystyle{\left[{\cal Q}_{{\bf q_1}}\!,
  {\cal Q}_{{\bf q_2}}^{+}\right]\!\!=\!
  e^{i\phi}\!\!{\cal A}_{\bf q_1\!-
  \!q_2}\!\!-\!e^{-i\phi}{\cal B}_{\bf q_1\!-
  \!q_2}} \eqno(\mbox{A.2})\vspace{-1mm}
$$
[$\phi\!=\!({{\bf q}_1}\!\times\!{{\bf q}_2})\!{}_z\!/2$], we obtain
$\left[{\hat{\cal H}}_0',{\cal Q}_{\bf q}^\dag\right]|0\rangle\!=\!(\epsilon_{\rm Z}\!+\!{\cal E}_q){\cal Q}_{\bf q}^\dag|0\rangle$, where the spin-wave Coulomb energy is
$\displaystyle{{\cal E}_q\!=\!\int_0^\infty\!\!\!\!pdp{\widetilde U}(\!p)[1\!-\!J_0(pq)]}$ (cf. Ref. \onlinecite{theory}).

Now with the help of Eqs. (A.1-2) one can calculate projection of one two-exciton state to
another:\vspace{1mm} $\langle 2;\{{\bf q}\}_2'|\{{\bf q}\}_2;2\rangle\!\equiv\!\left\langle{\cal Q}_{{{\bf q}_2}'}{\cal
Q}_{{{\bf q}_1}'}{\cal Q}_{{\bf q}_{\!1}}^{\dag}\!{\cal Q}_{{\bf q}_2}^{\dag}\right\rangle$\vspace{-3.mm}
$$\begin{array}{l}
\displaystyle{=\!{\cal N}_\phi^2\left(\!\delta_{{{\bf q}_2}'\!\!,{{\bf q}_2}}\delta_{{{\bf q}_1}'\!\!,{{\bf q}_1}}\!\!\!+\!\delta_{{{\bf q}_1}'\!\!,{{\bf q}_2}}\delta_{{{\bf q}_2}'\!\!,{{\bf q}_1}}\vphantom{\displaystyle{\left.-\!\frac{2\cos{\phi}}{{\cal N}_\phi}\delta_{{\bf q}_1\!+\!{\bf
q}_2,{{\bf q}_1}'\!\!+\!{{\bf q}_2}\!{}'}\!\!\right)}}\right.}\qquad\qquad\qquad\qquad\\
\qquad\qquad\qquad\qquad\displaystyle{\left.-\!\frac{2\cos{\!\Phi}}{{\cal N}_\phi}\delta_{{\bf q}_1\!+\!{\bf
q}_2,{{\bf q}_1}'\!\!+\!{{\bf q}_2}\!{}'}\!\!\right)}\!,\qquad ({\rm A.3})\vspace{-2mm}
\end{array}
$$
where $\Phi\!\!=\!\left({{\bf q}_1}'\!\!\times\!{{\bf q}_1}\!\!+\!{{\bf q}_2}'\!\!\times\!{{\bf
q}_2}\right)\!{}_z\!/{2}$. Were the ${\cal Q}$-operators simply Bose ones, then only the first two terms in the
parentheses of Eq. (A.3) would constitute the result of the four-operator expectation. However, the presence of the
third term is a manifestation of a `kinematic spin exciton interaction'. This `interaction' is a
consequence of the non-Bose commutation rules (A.2). Such a specific spin-excitonic `coupling' plays a role, for instance, in research of phenomena related to mutual spin exciton scattering, and also when calculating norms of
many-exciton states in case the total number of excitons is comparable to ${\cal N}_\phi$. However, if we find the
squared norm of two-exciton state $|2;\!\{{\bf q}\}_2\rangle$ with different momenta, ${\bf q}_1\!\neq\!{\bf q}_2$,
then the kinematic interaction may be neglected and we just have $\langle\{{\bf q}\}_2;2|2;\!\{{\bf
q}\}_2\rangle\!\approx\!{\cal N}_\phi^2$.
For the $|k;\!\{{\bf q}\}_k\rangle$ state at low exciton concentration, $k/{\cal N}_\phi\!\ll\!1$, considering all ${\bf q}$'s to be different, any interference of single spin-exciton states may be ignored.
It is quite sufficient for our calculations to use only the following recurrent property of the squared norm:\vspace{-1.mm}
$$\begin{array}{l}
\!\langle k;\{{\bf q}\}_k|\{{\bf q}\}_k;k\rangle\qquad\qquad\vspace{1mm}\\
=\left[{\cal N}_\phi+O(k)\right]\langle k\!-\!1;\{{\bf q}\}_{k\!-\!1}|\{{\bf q}\}_{k\!-\!1};k\!-\!1\rangle.\end{array} \eqno({\rm A}.4)
$$

State \eqref{state-q} represents a dilute gas of spin-wave excitons against the background of the Goldstone-exciton condensate. Let us act on it by the operator $S_+\!\equiv\!{\cal Q}_0$. Using properties
$$[S_+,{\cal
Q}_{\bf q}^\dag]\!=\!{\cal A}_{\bf q}^\dag\!-\!{\cal B}_{\bf q}^\dag\,,\quad [{\cal A}_0\!-\!{\cal B}_0,{\cal Q}_{\bf
q}^\dag]\!=\!-2{\cal Q}_{\bf q}^\dag\,,\vspace{-1mm}
$$
and \vspace{-2mm}
$$({\cal A}_{\bf q}^\dag\!-\!{\cal B}_{\bf q}^\dag)|0\rangle\!= \!{\cal
N}_\phi\delta_{{\bf q},0}|0\rangle,
$$
we come to the following equation:
$$
\!\!\!\!\begin{array}{l}
\!\!\!\!\!S_+|\{{\bf q}\}_k;n\rangle\!=\!(n\!-\!k)({\cal N}_\phi\!-\!n\!-\!k\!+\!1)|\{{\bf q}\}_k;n\!-\!1\rangle\vspace{.5mm}\\\!\!\!\!\!\!+\,\displaystyle{(S_-)^{n\!-\!k}\!\!\sum_{i=1}^{k\!-\!1}\!{\cal Q}_{{\bf q}_1}^\dag
{\cal Q}_{{\bf q}_2}^\dag...{\cal Q}_{{\bf q}_{i\!-\!1}}^\dag\!({\cal A}_{{\bf q}_i}^\dag\!\!-\!{\cal B}_{{\bf q}_i}^\dag)...{\cal Q}_{{\bf q}_k}^\dag|0\rangle}.\vspace{-3mm}
\end{array}\eqno({\rm A}.5)
$$
With the help of commutation rule (A.1) one finds that the second item in the r.h.s. here is determined by the
kinematic interaction of spin-wave excitons. We study the situation where $k\!\ll\!{\cal N}_\phi$ and, besides,
$k\!\ll\!{\cal N}_\phi\!-\!n$. Then the squared norm of the second item, being smaller than \vspace{-1mm}
$$
k^2\langle\{{\bf
q}\}_{k\!-\!1};n\!-\!1|n\!-\!1;\{{\bf q}\}_{k\!-\!1}\rangle,\vspace{-1mm}
$$
turns out to be negligible compared to that of the first
one, and, hence, in the r.h.s. of Eq. (A.4) we retain only the first term. By acting $n\!-\!k$ times with operator
$S_+$ on state $|\{{\bf q}\}\!{}_k;n\rangle$, we get\vspace{-1mm}
$$
{}\!{}\!(S_+)^{n\!-\!k}|\{{\bf q}\}\!{}_k;n\rangle\!\approx\!\displaystyle{\!\frac{(n\!-\!k)!({\cal N}_\phi\!-\!2k)!}{({\cal N}_\phi\!-\!n\!-\!k)!}|\{{\bf q}\}\!{}_k;k\rangle.} \eqno ({\rm A.6}) \vspace{-4mm}
$$
Then we come to the result given by equation \eqref{normk} for the squared norm of
state $|\{{\bf q}\}\!{}_k;n\rangle$.
In the special cases of $k\!=\!0$ and $k\!=\!1$, when the kinematic interaction is missing, formulae (A.6) and
\eqref{normk} are quite exact and result in Eqs. \eqref{norm_Nphi_n} and \eqref{norm_n_q}.

Finally, we note that in summation over ${\bf q}$ in Eq. \eqref{dk/dt}, one may certainly ignore any cases of exact
coincidence of the ${\bf q}\!=\!{\bf q}_{k\!+\!1}$ number in the final state $|\{{\bf q}\}\!{}_{k\!+\!1};n\rangle$ with
some of the values ${\bf q}_1$, ${\bf q}_2$,...${\bf q}_k$ in the initial one.  This is evident from the fact that the
zero-dimensional phase volume of these coincidences is negligible compared to the 1D volume of possible ${\bf q}$
values in the final state [the 1D volume, rather than the 2D one, is due to the presence of the $\delta$-function in Eq.
\eqref{dk/dt}]. This statement is true for any number $k$, and, therefore, only states with ${\bf q}_1\!\neq\!{\bf
q}_2\!\neq...\!\neq\!{\bf q}_k$ values are relevant in the framework of the kinetic approach developed in Sec. VI.

\vspace{-4mm}

\section{}

\vspace{-3mm}

The ${\bf q}\equiv 0$ equivalence corresponding to the Goldstone spin exciton $S_-|0\rangle\equiv {\cal
Q}^\dag_{\{{\bf q}\equiv 0\}}|0\rangle$ actually means $q\!\ll\! 1/{\mathscr L}$ [cf. inequality \eqref{photon}]  in
contrast to the `spin-orbital' excitation ${\cal Q}^\dag_{\{{\bf q}\to 0\}}|0\rangle$ { where $q\ll 1/l_B$ but}
$q\!\gg\!1/{\mathscr L}$ (here normal dimensionality of $q$ is used). In the ideally homogeneous system ${\mathscr
L}=L$ where $L$ is the 2D channel linear dimension. Indeed, in the presence of an external smooth random potential
(SRP) characterized by amplitude ($\Delta$) and correlation length ($\Lambda$), one definitely { assumes
${\mathscr L}$ to be at least not smaller than $\Lambda$,  because $1/\Lambda$ measures violation of the translation
invariance in the 2D system}. This condition is not the only one. The SRP lifts the Landau level degeneracy, and the
$|0\rangle$ orbital state is changed compared to the homogeneous case. In fact, the `standard' single-electron wave
function is localized in the 2D space near a `standard' equipotential line (EL), within a `belt' of width $l_B$ (see,
e.g., publication$\,$\cite{iord96}). The length of the closed standard EL corresponding to electron energy
$0<|\epsilon|\lesssim \Delta$ ($\epsilon$ is measured from the Landau level center) is of the order of $\Lambda$. If
${\mathscr T}$ is a classical period of drift motion in crossed fields ${\vec B}$ and ${\vec {\mathscr E}}$
($|{\vec {\mathscr E}}|\sim \Delta/\Lambda$) along the closed standard EL, then the level spacing between two adjacent
states is $\sim 1/{\mathscr T}\sim \Delta(l_B/\Lambda)^2$. So, taking into account the $e$-$e$ interaction, we
conclude that the ${\bf q}\equiv 0$ condition for a collective state means { that the Coulomb energy
${\cal E}_q$ becomes physically meaningless if it turns out to be smaller than} single-electron energy uncertainty
$1/{\mathscr T}$. That is, the formal condition ${\bf q}\equiv 0$ determining the undisturbed orbital state of the
quantum Hall ferromagnet means ${\cal E}_q\ll \Delta(l_B/\Lambda)^2$ resulting in $q\ll 1/{\mathscr L}\sim
(\Delta M_{\rm x})^{1/2}\!/\Lambda$. The {length ${\mathscr L}$} should be substituted into Eq. \eqref{photon}.
(In modern heterostructures $\Delta=0.5-0.7\,$meV and $\Lambda\simeq 50\,$nm.)

We note also that the essential difference between ${\bf q}\!\to\!0$ and ${\bf q}\!\equiv\!0$ spin excitations
studied in this paper is not a feature peculiar only to a quantum Hall ferromagnet. Just the same situation takes place
in the case of a dielectric ferromagnet representing,for instance, a system of atomic spins spatially localized at
crystal-lattice sites and described by the Heisenberg model. The Bloch operator creating a magnon with wave vector
${\bf k}$ is $\hat{\mathscr S}_{\bf k}\propto \sum_{\bf n}\!e^{\mbox{i}{\bf k}{\bf r_n}}S_{{\bf n}-}$, where ${\bf
r_n}$ is the lattice site position and $S_{{\bf n}-}$ is the spin-lowering operator acting on the spin in the ${\bf
n}$-th site.\cite{LP} This operator is very similar to spin-wave operator \eqref{QQ} if the $q_y\!=\!0$ condition
holds (which may be always ensured by simply choosing the ${\hat x}$ axis directed along momentum ${\bf q}$).
Besides, it also reduces to total spin-lowering operator $S_-\!=\!\sum_{\bf n}\!S_{{\bf n}-}$ if ${\bf
k}\!\equiv\!0$. The state $\hat{\mathscr S}_{\bf k}|0\rangle$ (where in the ground state $|0\rangle$ all spins are
strictly polarized) is an eigenstate of the Heisenberg Hamiltonian and simultaneously an eigenstate for
total operators ${\hat S}^2$ and ${\hat S}_z$.\cite{LP} Routinely calculating quantum numbers $S$ and $S_z$, one can
see that the magnon state has spin numbers changed by $\delta S_z\!=\!-1$ and by $\delta S\!=\delta_{{\bf k},{\bf 0}}-1$ as
compared to the ground state.

\vspace{-6mm}

\section{}

\vspace{-2mm}

The summation in Eq. \eqref{dk/dt}, $\sum_{\bf q}\!...\delta(q^2\!/2M_{\rm x})$ where ${\bf q}$ values,  even
when infinitely small, are not identically zero, formally results in zero. However, if one adds an infinitesimal term
$({\bf q}\!\times\!\mbox{\boldmath $\varepsilon$})_z$ to the $\delta$-function argument ($\varepsilon\to 0$), then the
situation becomes well defined. The physical meaning of this term will become clear if one takes into account
the existence of electric dipole moment $el_B{\bf q}\!\times\!\!{\hat z}$ of the spin-wave exciton. That is, the 2D
vector $\mbox{\boldmath $\varepsilon$}$ is just proportional to a weak external electric field $\vec{\mathscr E}(x,y)$
appearing, for instance, due to a smooth random potential present in the quantum well. Thu,s the summation is performed
trivially\vspace{-3mm}
$$
\begin{array}{l}
\displaystyle{\sum_{\bf q}F({\bf q})\delta(...)}\qquad\qquad\qquad\qquad\qquad\qquad\qquad\vspace{-4mm}\\
\qquad\displaystyle{=({\cal N}_\phi/{2\pi})\lim_{\mbox{\boldmath $\varepsilon$}\!\to 0}\!\int\!\!d{\bf q}\,F({\bf q})\delta(q^2\!/2M_{\rm x}\!+{\bf q}\!\times\!\mbox{\boldmath $\varepsilon$})}\vspace{1mm}\\
\qquad\qquad\qquad\qquad\qquad\qquad\qquad\displaystyle{=F(0){\cal N}_\phi M_{\rm x}\!/2.}
\end{array}\vspace{-5mm}
$$


\begin{thebibliography}{99}
\bibitem{con-spin-rot}
As applied to the $\nu\!=\!1$ quantum Hall ferromagnet this situation was considered;\cite{by99} in this case the coefficients are $C_n\!=\![\cos{\!(\!{\beta}/{2})}]^{{{\cal N}_\phi}\!\!-n}
[-\sin{\!({\beta}\!/{2})}]^n\!\!/{n!}$ [cf. Eq. (3.12) at $N\!=\!{\cal N}_\phi$].
\bibitem{foot0}
{Most of these recombination processes$\,$\cite{Fukuoka,la15} is a `gemini' recombination where an electron--heavy-hole
pair born due to photon absorption annihilates before both quasi-particles have had sufficient time to move apart at a
distance greater than the spatial scale of their wave functions. However,  there is also a recombination channel where
the born heavy hole recombines not with its `twin' electron but due to  `vertical' (radiative) transition with a
conduction-band electron located at the same point in space. Then the twin electron, `inclined' by angle $\beta$ in the
spin space owing to total angular momentum conservation, occupies the vacant site occuring in the Landau level in
the conduction band and thus resulting in $\!\uparrow\,\to\!{\nearrow}$ replacement. The recombination transitions are
accopanied by appropriate irradiation processes and by fast spin relaxation ocurring within the valence band.}
\end{thebibliography}
\end{document}